\documentclass[fleqn]{aa}  

\usepackage{graphicx}
\usepackage{xcolor}
\usepackage{hyperref}
\usepackage{txfonts}
\usepackage{calligra}
\usepackage{calrsfs}
\usepackage[T1]{fontenc}
\usepackage[mathcal]{eucal}
\usepackage{longtable, caption}
\usepackage{orcidlink}

\begin{document}

\title{Flashlights: Prospects for constraining the initial mass function around cosmic noon with caustic-crossing events}

\titlerunning{Flashlights: constraining stellar IMF}

\author{Ashish Kumar Meena\inst{\orcidlink{0000-0002-7876-4321},1,\thanks{ashishmeena766@gmail.com}}
    \and Sung Kei Li\inst{\orcidlink{0000-0002-4490-7304},2}
    \and Adi Zitrin\inst{\orcidlink{0000-0002-0350-4488},1}
    \and Patrick L. Kelly\inst{\orcidlink{0000-0003-3142-997X},3}
    \and Tom Broadhurst\inst{\orcidlink{0000-0002-8785-8979},4,5,6}
    \and Wenlei Chen\inst{\orcidlink{0000-0003-1060-0723},7}
    \and Jose M. Diego\inst{\orcidlink{0000-0001-9065-3926},8}
    \and Alexei V. Filippenko\inst{\orcidlink{0000-0003-3460-0103},9}
    \and Lukas J. Furtak\inst{\orcidlink{0000-0001-6278-032X},1}
    \and Liliya L. R. Williams\inst{\orcidlink{0000-0002-6039-8706},3}}

\institute{Department of Physics, Ben-Gurion University of the Negev, PO Box 653, Beer-Sheva 84105, Israel
    \and Department of Physics, The University of Hong Kong, Pokfulam Road, Hong Kong
    \and School of Physics and Astronomy, University of Minnesota, 116 Church Street SE, Minneapolis, MN 55455, USA
    \and Donostia International Physics Center, DIPC, Basque Country, San Sebastián, 20018, Spain
    \and Department of Physics, University of Basque Country UPV/EHU, Bilbao, Spain
    \and Ikerbasque, Basque Foundation for Science, Bilbao, Spain
    \and Department of Physics, Oklahoma State University, 145 Physical Sciences Bldg, Stillwater, OK 74078, USA
    \and IFCA, Instituto de F´ısica de Cantabria (UC-CSIC), Av. de Los Castros s/n, 39005 Santander, Spain
    \and Department of Astronomy, University of California, Berkeley, CA 94720-3411, USA}

\abstract{The Flashlights program with the \textit{Hubble} Space Telescope imaged the six Hubble Frontier Fields galaxy clusters in two epochs and detected twenty transients. These are primarily expected to be caustic-crossing events (CCEs) where bright stars in distant lensed galaxies, typically at redshift $z\approx1$--3, get temporarily magnified close to cluster caustics. Since CCEs are generally biased toward more massive and luminous stars, they offer a unique route for probing the high end of the stellar mass function. We take advantage of the Flashlights event statistics to place preliminary constraints on the stellar initial mass function (IMF) around cosmic noon. The photometry (along with spectral information) of lensed arcs is used to infer their various stellar properties, and stellar synthesis models are used to evolve a recent stellar population in them. We estimate the microlens surface density near each arc and, together with existing lens models and simple formalism for CCEs, calculate the expected rate for a given IMF. We find that, on average, a Salpeter-like IMF ($\alpha=2.35$) underpredicts the number of observed CCEs by a factor of ${\sim}0.7$, and a top-heavy IMF ($\alpha=1.00$) overpredicts by a factor of ${\sim}1.7$, suggesting that the average IMF slope may lie somewhere in between. However, given the large uncertainties associated with estimating the stellar populations, these results are strongly model-dependent. Nevertheless, we introduce a useful framework for constraining the IMF using CCEs. Observations with \textit{James Webb} Space Telescope are already yielding many more CCEs and will soon enable more stringent constraints on the IMF at a range of redshifts.}


\maketitle

\section{Introduction}
\label{sec:intro}
Understanding the formation and evolution of stars and galaxies across cosmic history is one of the main interests in astronomy~\citep[e.g.,][]{2014ARA&A..52..291C, 2014ARA&A..52..415M}. Giant, supergiant, and hypergiant stars (which we shall call throughout ``supergiants" for simplicity) dominate the light coming from young and star-forming galaxies. Such stars only live for several million years~(Myr) and end their life through supernovae, further affecting the properties of their host galaxies~\citep[e.g.,][]{2012MNRAS.419..479E, 2022ARA&A..60..455E}. In general, we can only observe and study individual supergiant stars directly within our own Galaxy or its vicinity~\citep[e.g.,][]{2007ApJ...659.1198E, 2018ApJ...868...57C}, while for the more distant universe~(e.g., redshift $z\gtrsim1$), we rely on integrated light or comparison with synthesized stellar population models to constrain their properties. Our understanding of the formation and evolution of supergiant stars and their effects on the integrated light of the host galaxy is thus still far from complete~\citep[e.g.,][]{2013ARA&A..51..393C, 2017PASA...34...58E}. One of the key missing pieces of the puzzle is their abundance; for example, various studies suggest that there may be significantly more blue supergiants than predicted by classical stellar evolution theory (sometimes referred to as the ``blue supergiant problem"; e.g.,~\citealt{2024ApJ...967L..39B}).

Given that supergiants are short-lived and thus probe recent phases of star formation, their abundance in different galaxies through cosmic history can also help assess the universality -- or evolution -- of the stellar initial mass function (IMF), especially on its high-mass end, which would in turn have strong implications for galaxy evolution models. This may have become even more timely now, in light of recent results from \textit{James Webb} Space Telescope~(JWST). For example, some JWST results suggest an overabundance of ultraviolet (UV)-bright galaxies at cosmic dawn \citep[e.g.,][]{2024MNRAS.531.2615C, McLeod2024MNRAS.527.5004M}, and sufficient contribution by faint galaxies to the reionization of the Universe \citep[][c.f. \citealt{Naidu2020ApJ...892..109N}]{2024Natur.626..975A}, where most of the UV radiation in them is believed to come from hot, massive stars. Some early galaxies have been found to show hints of early mass build-up \citep{Labbe2023Natur.616..266L, 2024NatAs...8..657B}, which may be explained by very efficient star-formation channels \citep[e.g.,][]{Dekel2023MNRAS.523.3201D} or top-heavier IMFs \citep[e.g.,][]{Steinhardt2022ApJ...934...22S}.

On average, galaxies decrease in size and brightness with redshift~\citep[e.g.,][]{2004ApJ...611L...1B, 2015ApJ...808....6H, 2015ApJ...804..103K, 2019ApJ...880...57M, 2024MNRAS.527.6110O, 2023MNRAS.518.6011D}, but thanks to gravitational lensing magnification, some background galaxies, especially behind galaxy clusters, can get magnified substantially, revealing small-scale details within them~\citep[e.g.,][]{2014ApJ...790..143W, 2018AJ....155..104R, 2024Natur.632..513A, Fujimoto2024arXiv240218543Grapes}. A few years ago, lensing opened a route to observing individual stars at cosmological distances~\citep{2018NatAs...2..334K}. When stars in strongly magnified background galaxies that sit atop (or near) the lensing critical curves of a galaxy cluster lens, where the magnification gets extremely high, sweep across the net of microcaustics created by the microlenses in the intracluster medium~(ICM), they get sufficiently magnified to be observed for short periods of time. Such events are known as caustic transients or caustic-crossing events~\citep[CCEs;][]{MiraldaEscude1991ApJ...379...94M,2017ApJ...850...49V, 2018ApJ...857...25D, 2018PhRvD..97b3518O}. Roughly 80--90~(confirmed and candidates) of these CCEs have by now been detected with the \emph{Hubble} Space Telescope~(HST) and JWST, at various redshifts all the way to the first billion years in the Universe~\citep[e.g.,][]{2018NatAs...2..324R, 2019ApJ...881....8C, 2019ApJ...880...58K, 2022A&A...665A.134D, 2022arXiv221102670K, 2023A&A...672A...3D, 2023MNRAS.521.5224M, 2023ApJ...944L...6M, 2023ApJS..269...43Y, 2024A&A...681A.124D, 2024arXiv240408045F} with numbers continuously increasing. In addition to observing these CCE events in HST and JWST images, spectra have been obtained of some of these lensed star candidates~\citep{2022A&A...665A.134D, 2024MNRAS.527L...7F, 2024ApJ...976..166P, 2024arXiv240506953C}.

The peak magnification a star obtains when crossing a microcaustic depends on various factors such as the radius of the star, underlying (macro-) magnification, and the microlens density along with its mass function~\citep[e.g.,][]{2017ApJ...850...49V, 2018PhRvD..97b3518O}. High-cadence observations around these peaks can even reveal intrinsic properties of the background individual stars~\citep[e.g.,][]{2024ApJ...964..160H}. CCEs are expected mostly close to the macrocritical curves~(usually $\lesssim0.3''$ but can go up to~$\sim1''$ for low-redshift lensed arcs such as the Dragon arc in Abell~370 galaxy cluster;~\citealt{2024arXiv240408045F}), where the underlying macromagnification is sufficiently high, on the order of~$\gtrsim10^{2}$. However, very close to the macrocritical curve, there forms a so-called corrugated network of microcaustics such that the peak magnification remains on average constant, and lensed star candidates lying within it can also be observed as (semi-) persistent sources that remain observable for longer periods of time~\citep[e.g.,][]{2022Natur.603..815W, 2022ApJ...940L...1W,2023ApJ...944L...6M}. Farther away from the macrocritical curve, the macromagnification decreases~(as~$\mu_{\rm m} \propto 1/d$, where~$d$ is the distance from the macrocritical curve) along with the probability of observing CCEs. Nevertheless, the smaller the magnification, the larger the area, and the stronger also is the bias toward observing more luminous stars as they require less magnification to be observed. Hence, CCEs are an excellent route for tracing and for constraining the number density of this intrinsically luminous population of stars in the distant Universe and, thus, for constraining their IMF~\citep{Windhorst2018ApJS..234...41W}. In addition, since the event rate depends on the mass function and surface density of microlenses, CCEs also constitute a useful means for understanding the composition of dark matter in the intracluster medium \citep[ICM; e.g.,][]{2018ApJ...867...24D, 2018ApJ...857...25D, 2018PhRvD..97b3518O, 2024A&A...689A.167D, 2025ApJ...978L...5B, 2025MNRAS.536.1579M}.

In our current work, we use the event statistics of CCEs detected in Flashlights -- an HST program designed to detect CCEs and other transients~\citep{2022arXiv221102670K} -- to put preliminary constraints on the IMF in lensed galaxies around cosmic noon~($z\approx1$--3). Given that this is a relatively new field and the number of CCEs detected by Flashlights is relatively small, we study here the effect of two different IMFs~(parameterized as ~$dn/dm\propto m^{-\alpha}$), top-heavy~($\alpha=1.0$) and Salpeter~($\alpha=2.35$; \citealt{1955ApJ...121..161S}), on the expected number of CCEs. We focus on seventeen different lensed arcs, seven of those lead to the twenty CCEs observed in Flashlights as reported by~\citet{2022arXiv221102670K}. In addition to presenting the formalism to calculate the number of CCEs, we discuss various sources of uncertainty. Future work, including more events (which also helps reduce the related uncertainties), could then use a similar framework to put more stringent constraints on the IMF.

The paper is organized as follows. In Section \ref{sec:data}, we briefly overview the data used in this work along with the Flashlights program. Section \ref{sec:rate} discusses the methodology to estimate the CCE rate. We estimate, in Section~\ref{ssec:ssps_bound}, the number of CCEs expected in Flashlights and confront it with the observations. Section~\ref{ssec:lum_fun} studies the prospects for constraining the luminosity function of the background stellar population directly from the observed CCEs. Discussion regarding various sources of uncertainties is presented in Section~\ref{sec:uncertain}. We conclude in Section \ref{sec:conclusions}. Throughout this work, we use H$_0=70\:{\rm km\:s^{-1}\:Mpc^{-1}}$, $\Omega_{m}=0.3$, and $\Omega_{\Lambda}=0.7$. All magnitudes are given in the AB system~\citep{1983ApJ...266..713O}.

\section{Data and observations}
\label{sec:data}
We used very deep imaging of the six~Hubble Frontier Fields~\citep[HFF;][]{2017ApJ...837...97L}\footnote{\url{https://archive.stsci.edu/prepds/frontier/}} clusters in the ultrawide F200LP and F350LP WFC3/UVIS filters acquired under the Flashlights program~\citep[PI: Patrick L. Kelly;][]{2022arXiv221102670K}. Since two of the main goals of the Flashlights program were to understand the constituents of dark matter in the ICM and probe the stellar population at~$z\approx1$--3 using caustic-crossing~(and other) transients, each of these clusters was observed in two different epochs, with a 5$\sigma$ depth of~${\sim}29.3$~AB per epoch. As discussed by~\citet{2022arXiv221102670K}, Flashlights has led to the observation of twenty transients in the HFF clusters, including one off-caustic transient at~$z>1$ and a distance of~$\sim0.64''$ from the critical curve (where the macromagnification factor is small,~${\sim45}$), discussed by~\citet{2023MNRAS.521.5224M}. We refer readers to~\citet{2022arXiv221102670K} for more details about the Flashlights program.

All of the HFF clusters were earlier also observed under the Frontier Fields Survey program in three HST-ACS/WFC~(F435W, F606W, and F814W) and four HST-WFC3/IR~(F105W, F125W, F140W, and F160W) filters, complementing previous observations and reaching a~$5\sigma$ limiting magnitude of~${\sim}29$~AB~\citep{2017ApJ...837...97L}. We used these observations to measure the photometry and derive relevant stellar properties for each arc, and to estimate the microlens density in the vicinity of the arcs. For these purposes we used the \textsc{Bayesian Analysis of Galaxies for Physical Inference and Parameter EStimation}~\citep[\textsc{bagpipes};][]{2018MNRAS.480.4379C} and \textsc{spisea}~\citep{2020AJ....160..143H} codes. More details are given in the section below.

\begin{figure*}[!ht]
    \centering
    \includegraphics[width=4.0cm,height=4.0cm]{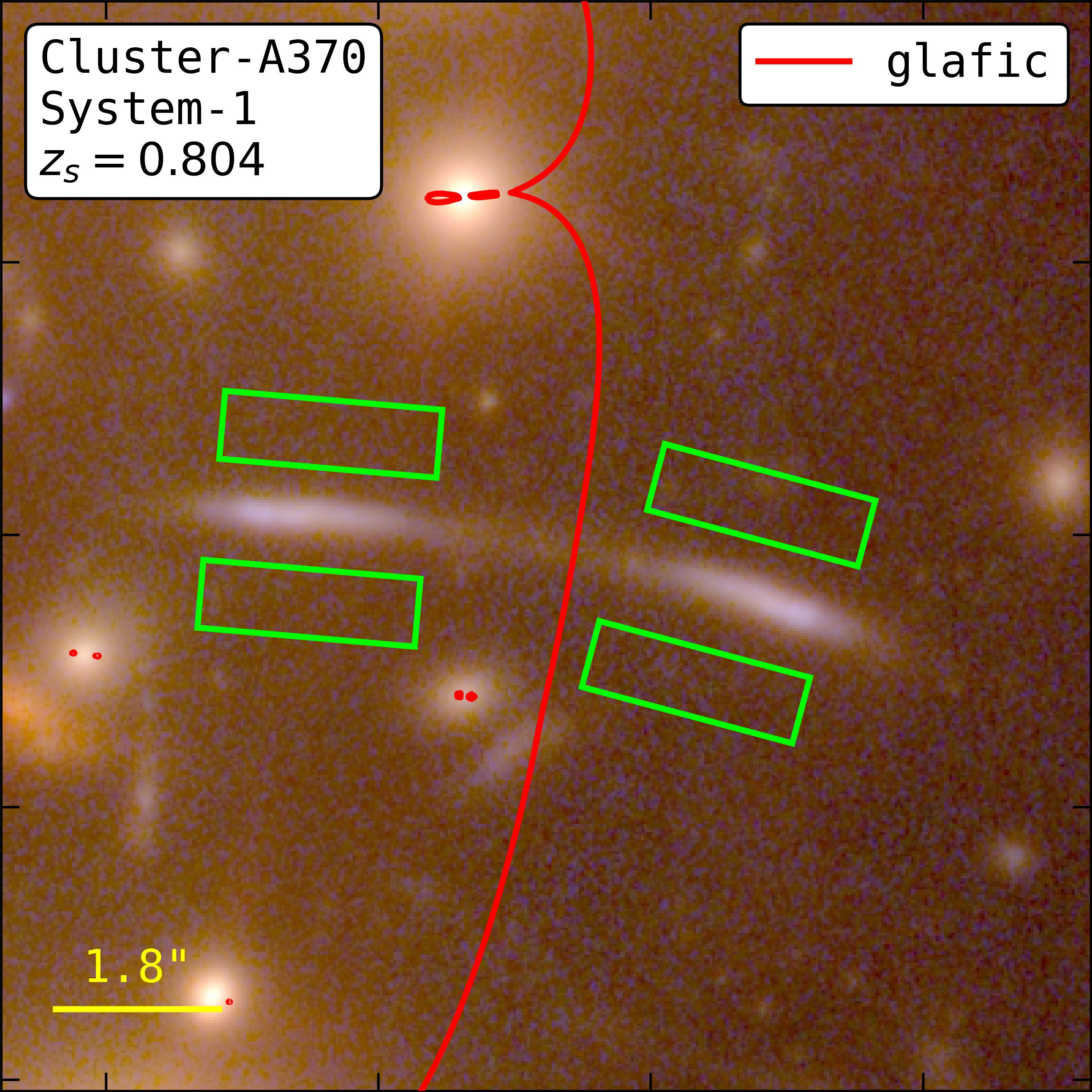}
    \includegraphics[width=4.0cm,height=4.0cm]{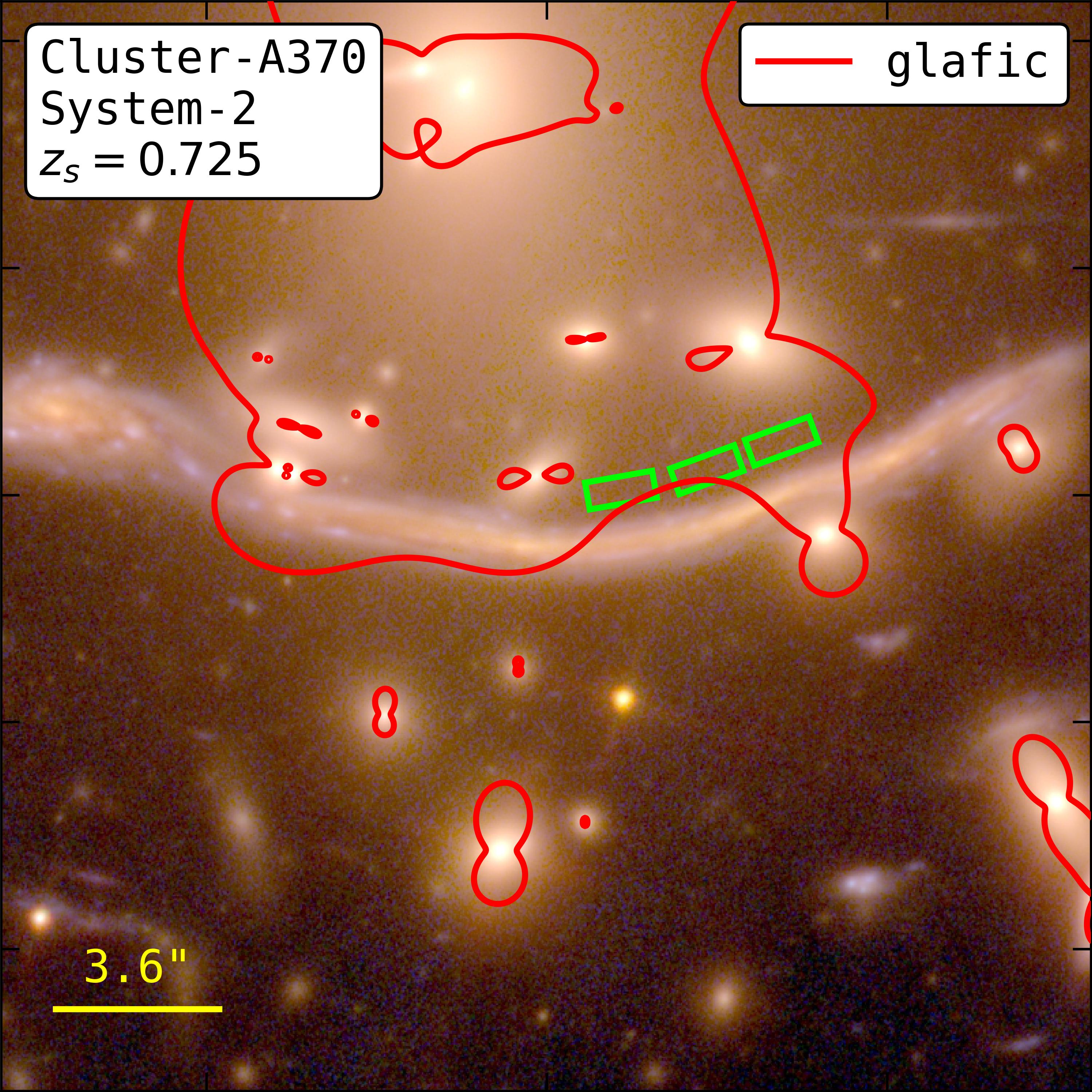}
    \includegraphics[width=4.0cm,height=4.0cm]{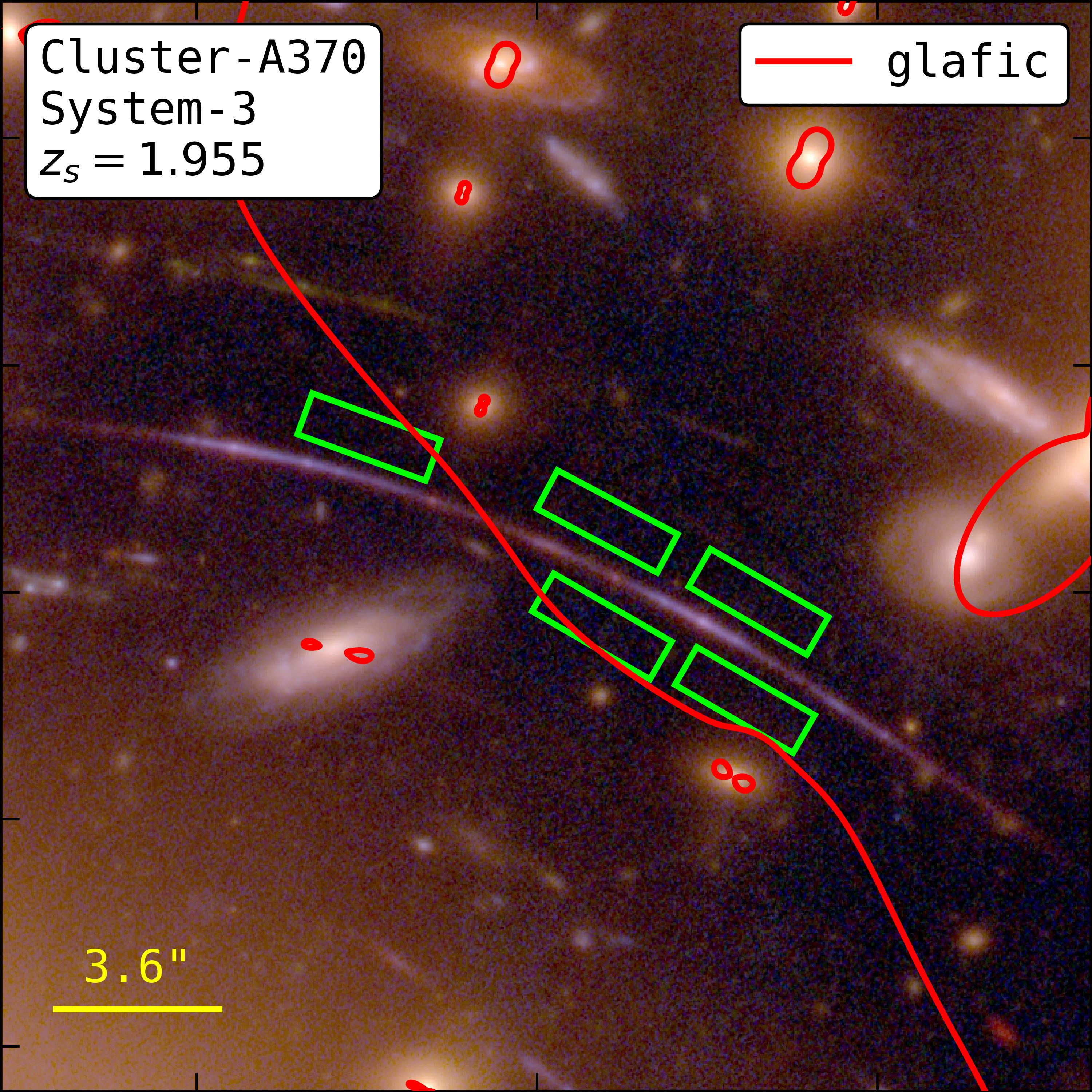}
    \includegraphics[width=4.0cm,height=4.0cm]{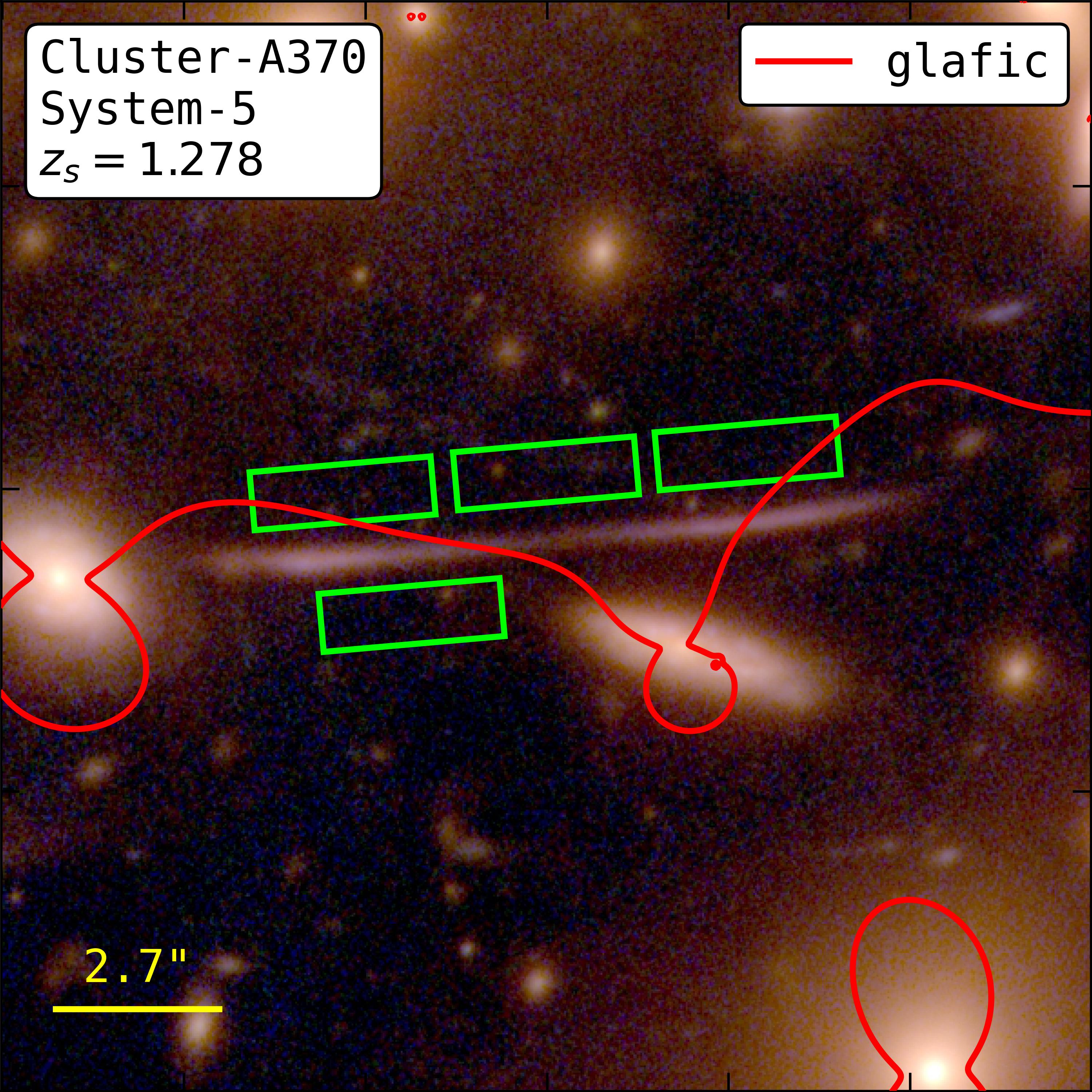}
    \includegraphics[width=4.0cm,height=4.0cm]{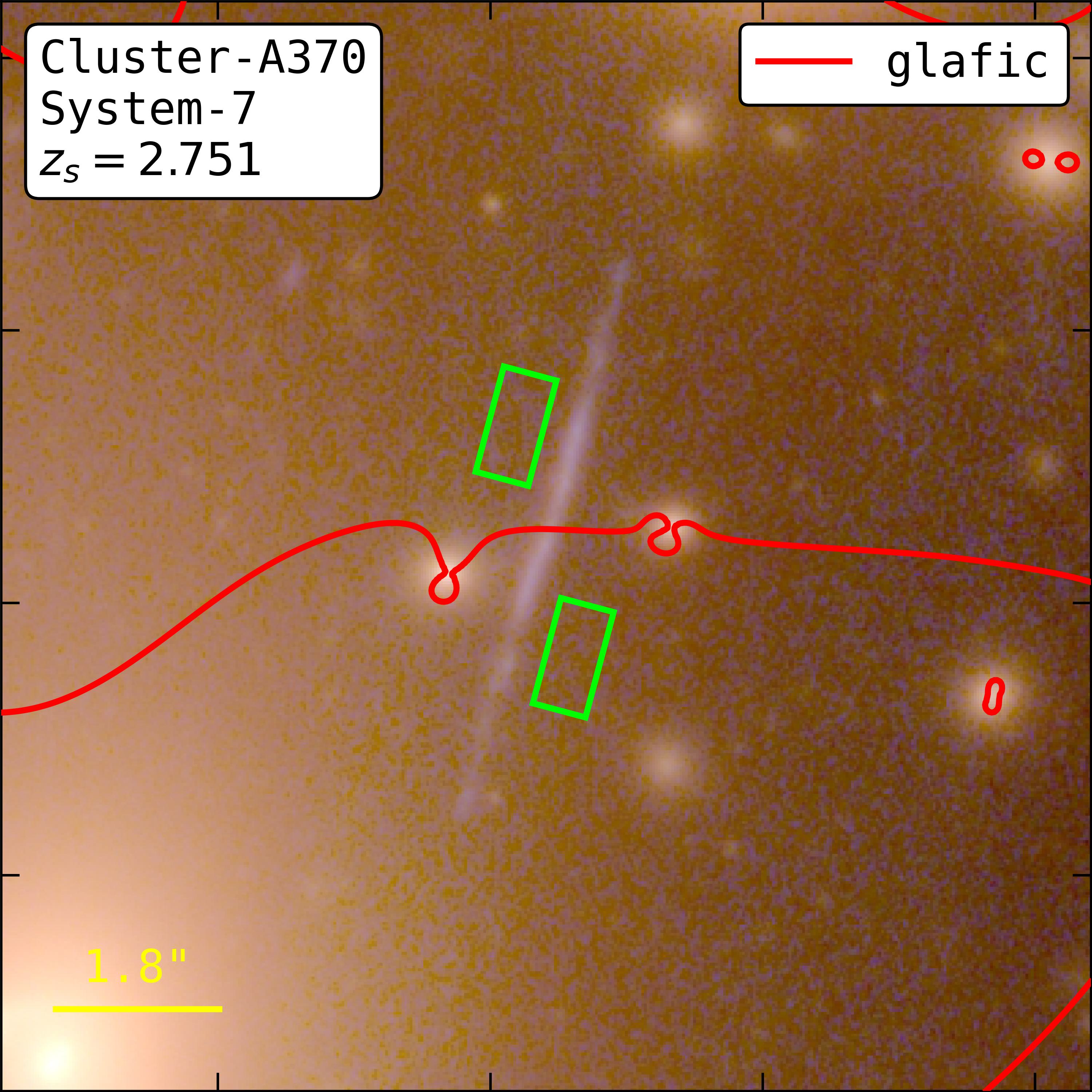}
    \includegraphics[width=4.0cm,height=4.0cm]{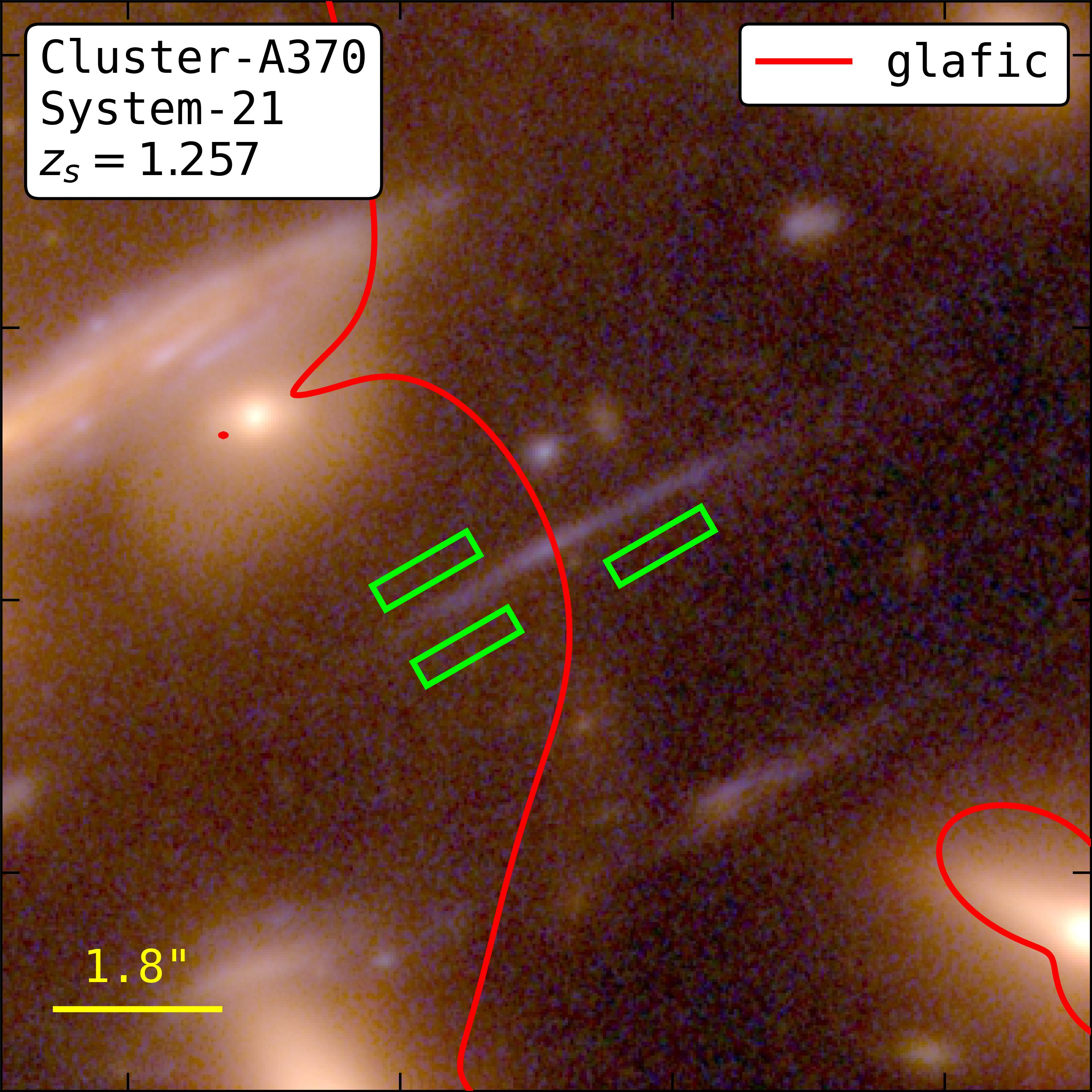}
    \includegraphics[width=4.0cm,height=4.0cm]{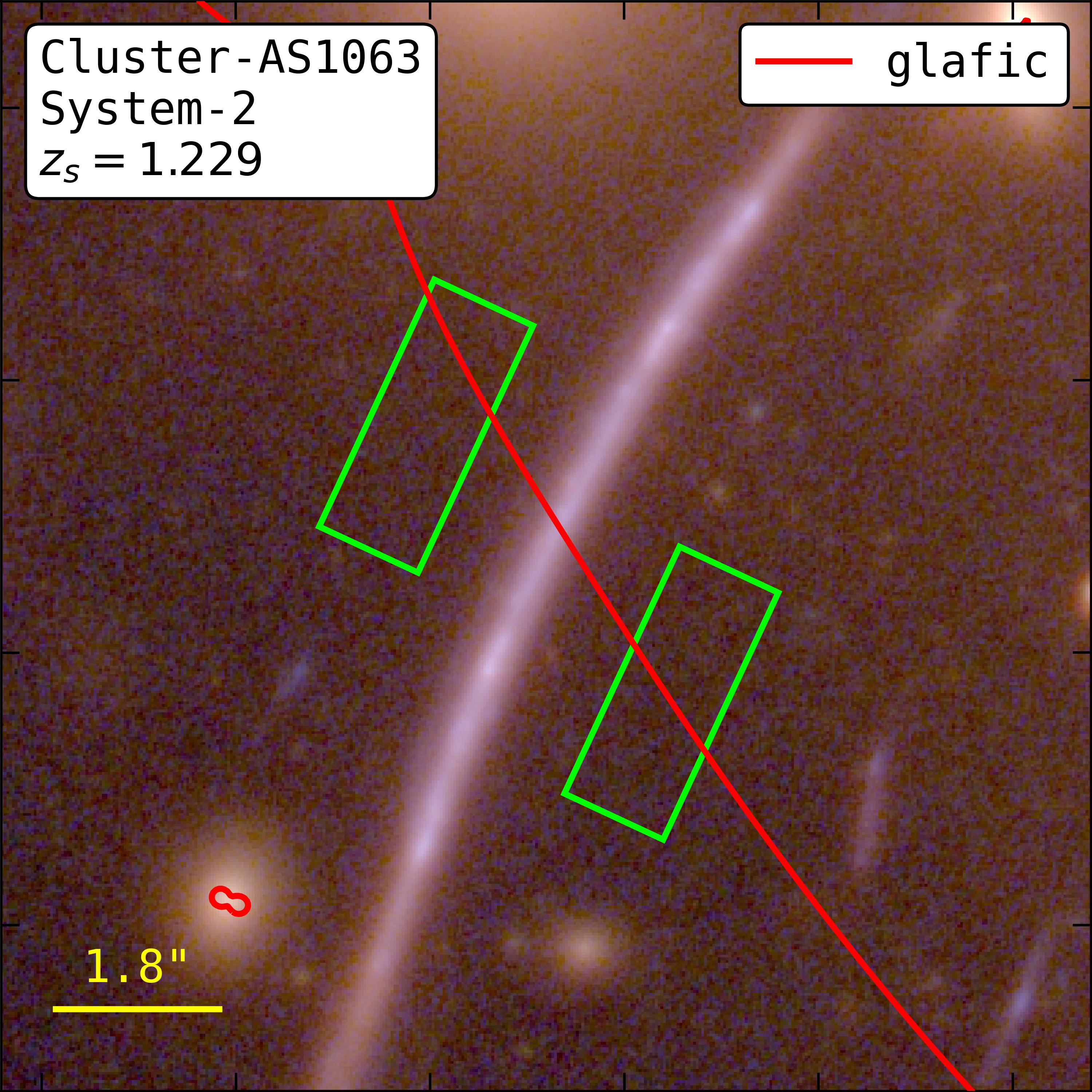}
    \includegraphics[width=4.0cm,height=4.0cm]{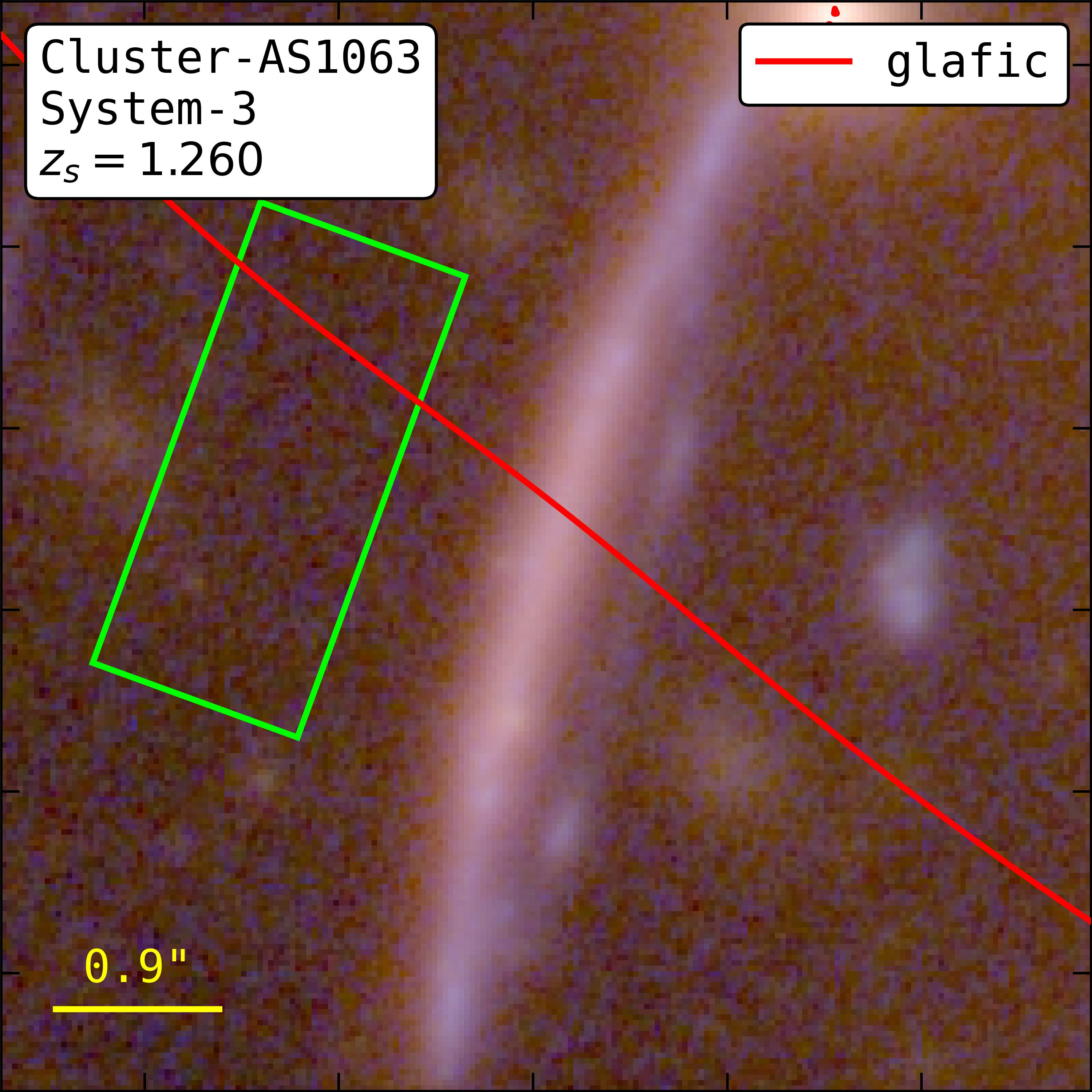}
    \includegraphics[width=4.0cm,height=4.0cm]{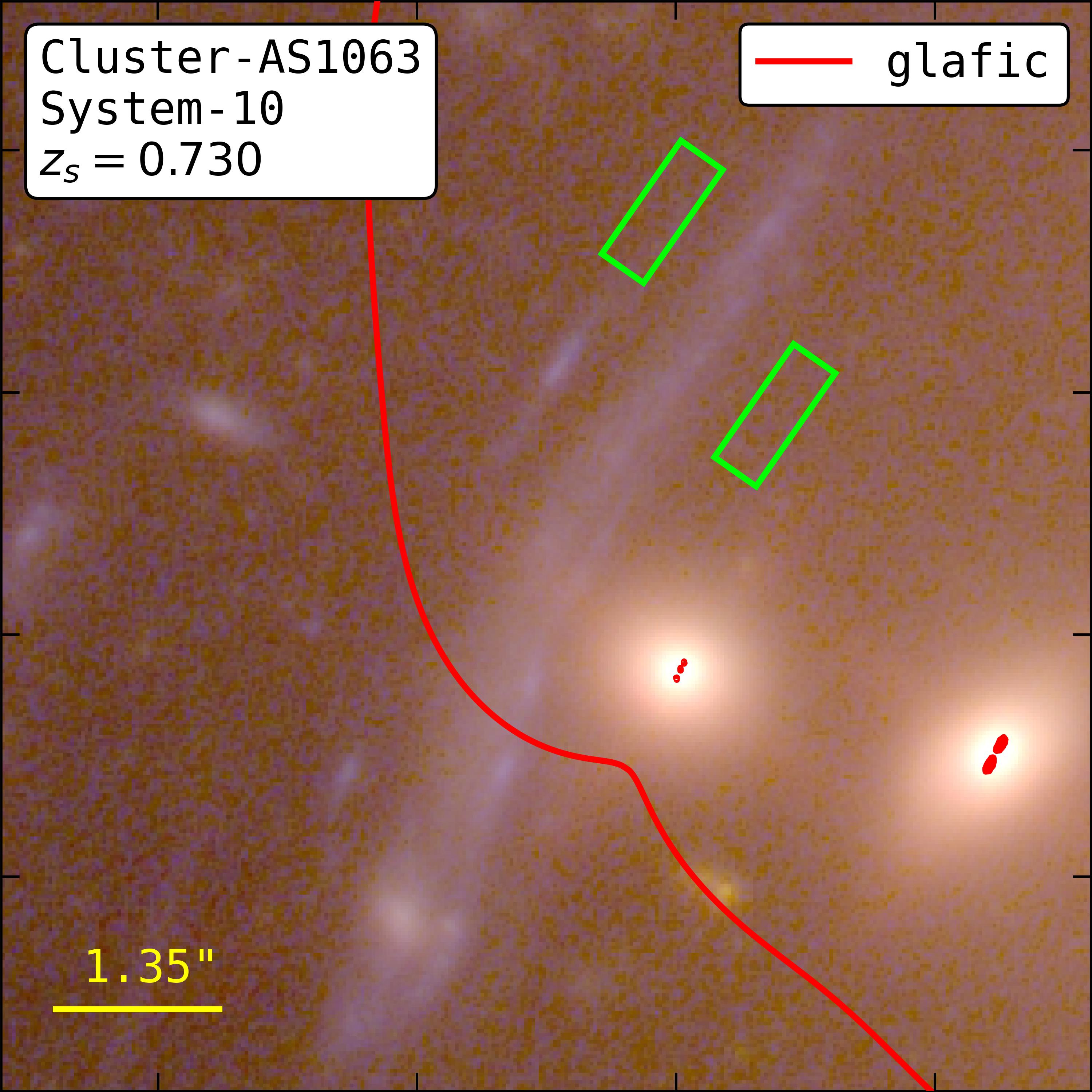}
    \includegraphics[width=4.0cm,height=4.0cm]{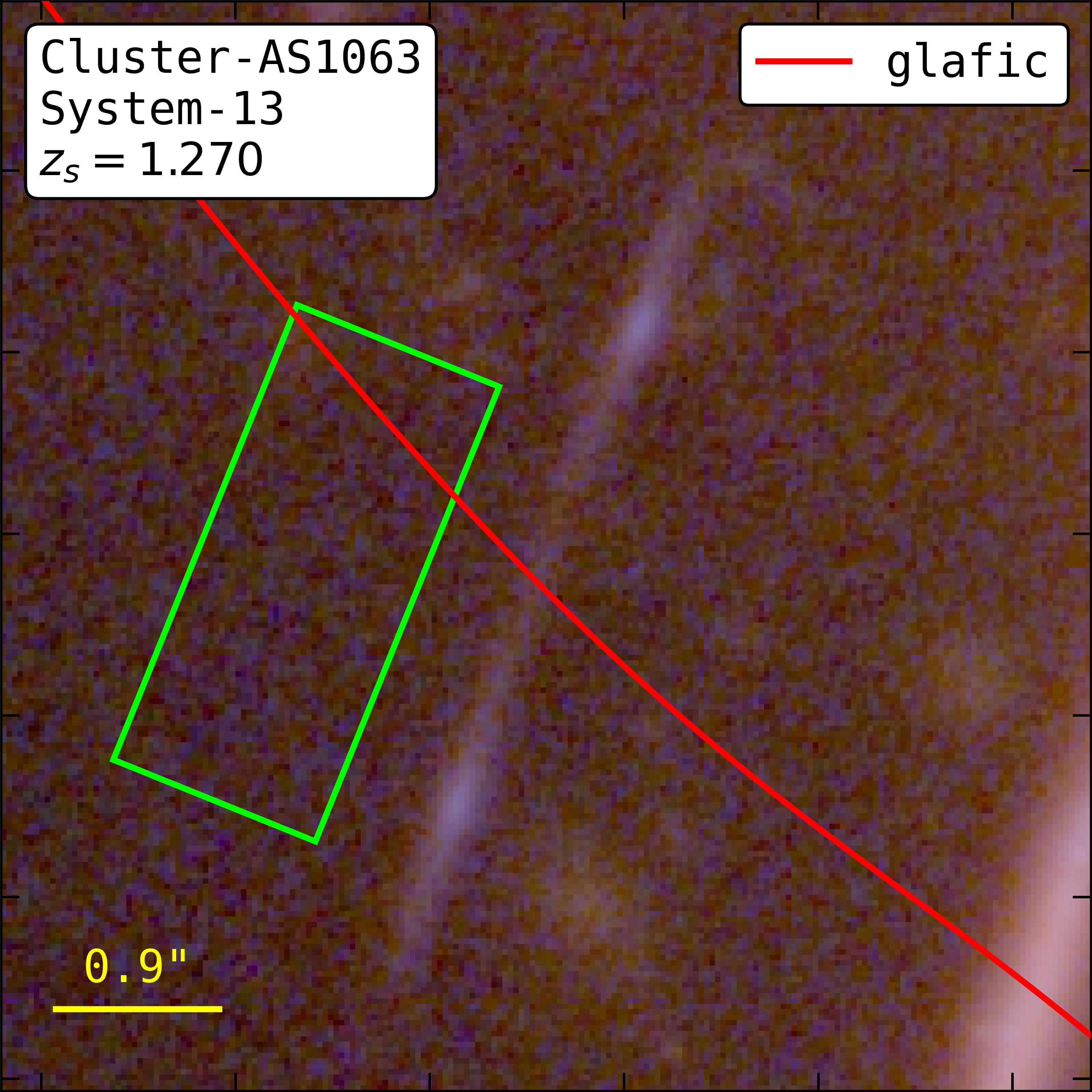}
    \includegraphics[width=4.0cm,height=4.0cm]{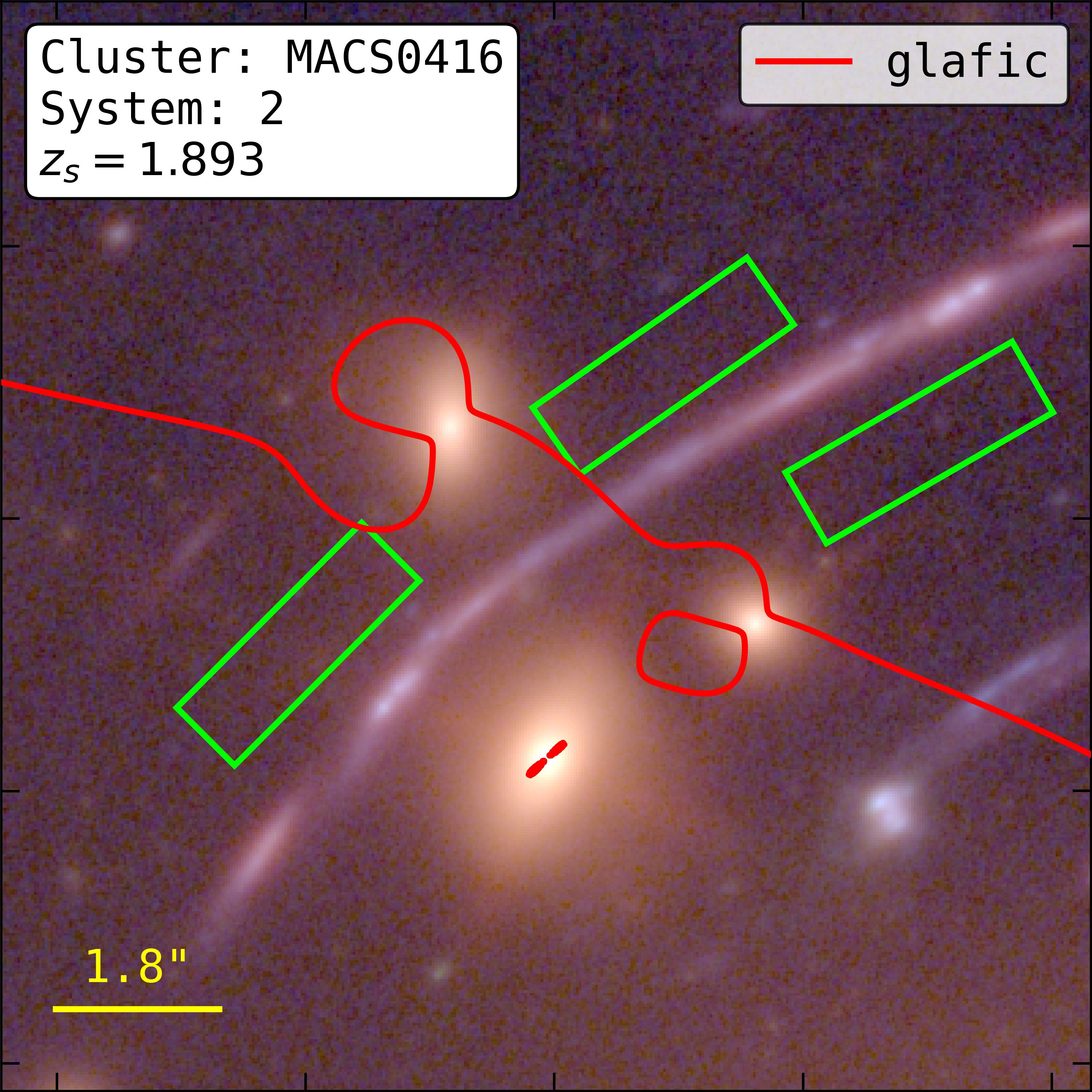}
    \includegraphics[width=4.0cm,height=4.0cm]{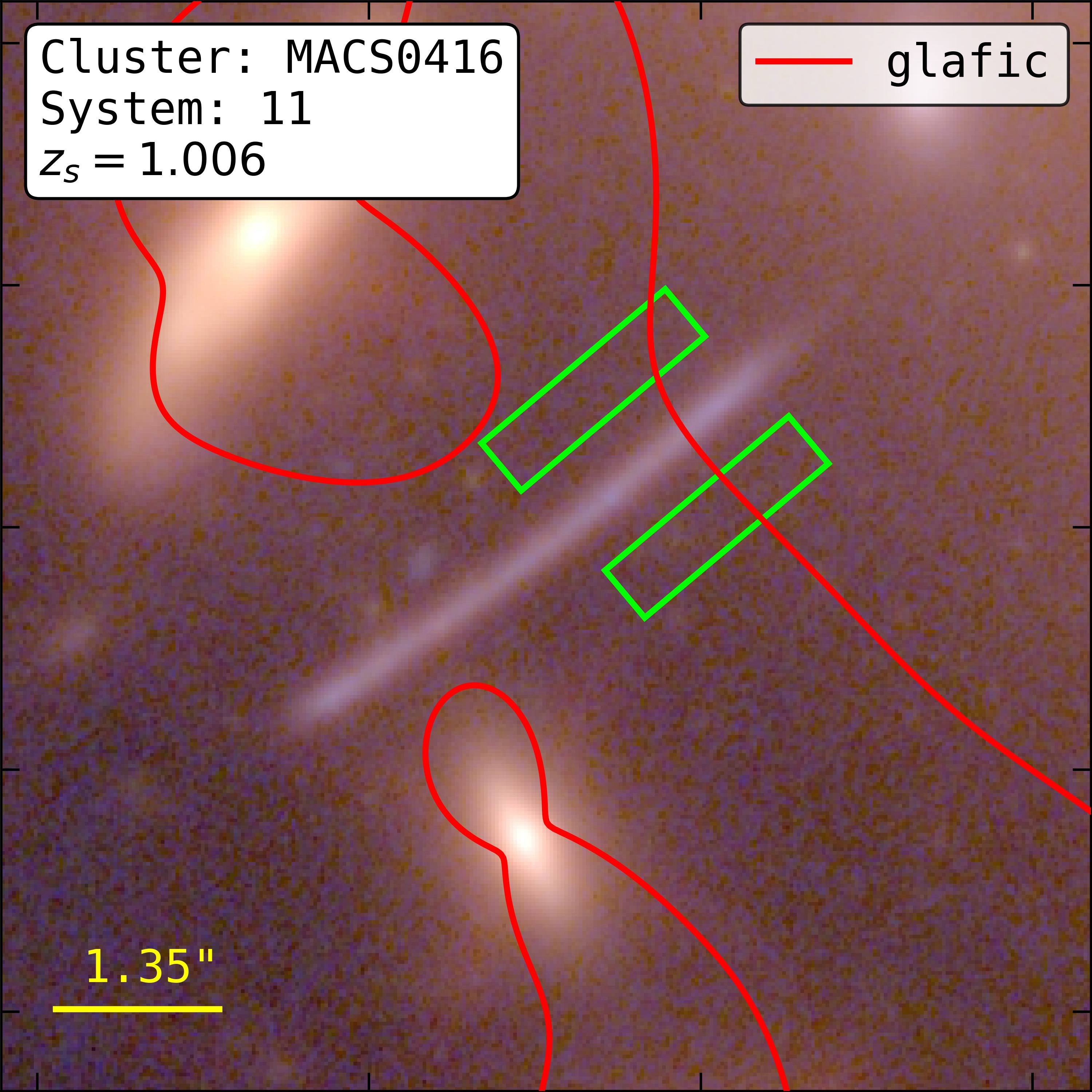}
    \includegraphics[width=4.0cm,height=4.0cm]{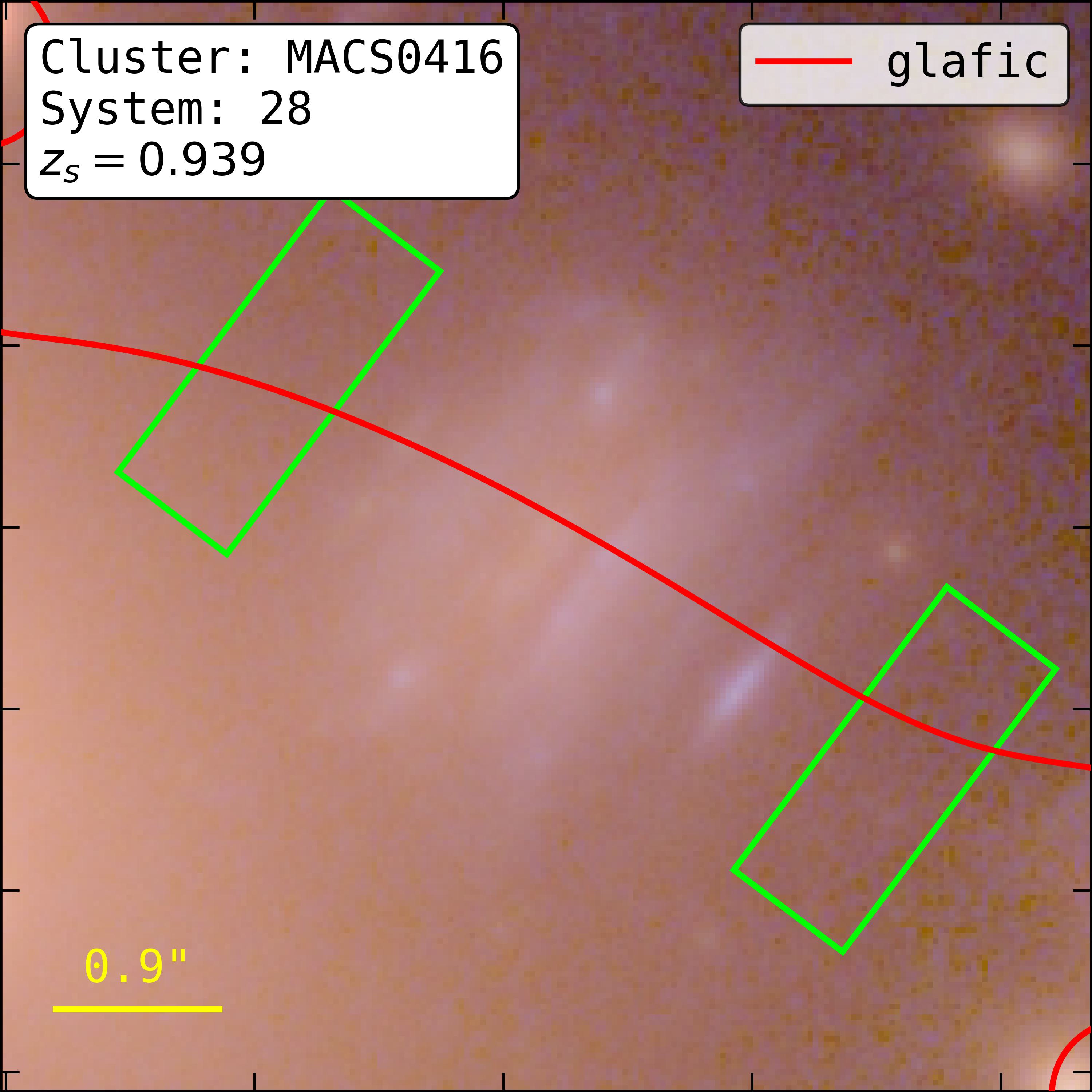}
    \includegraphics[width=4.0cm,height=4.0cm]{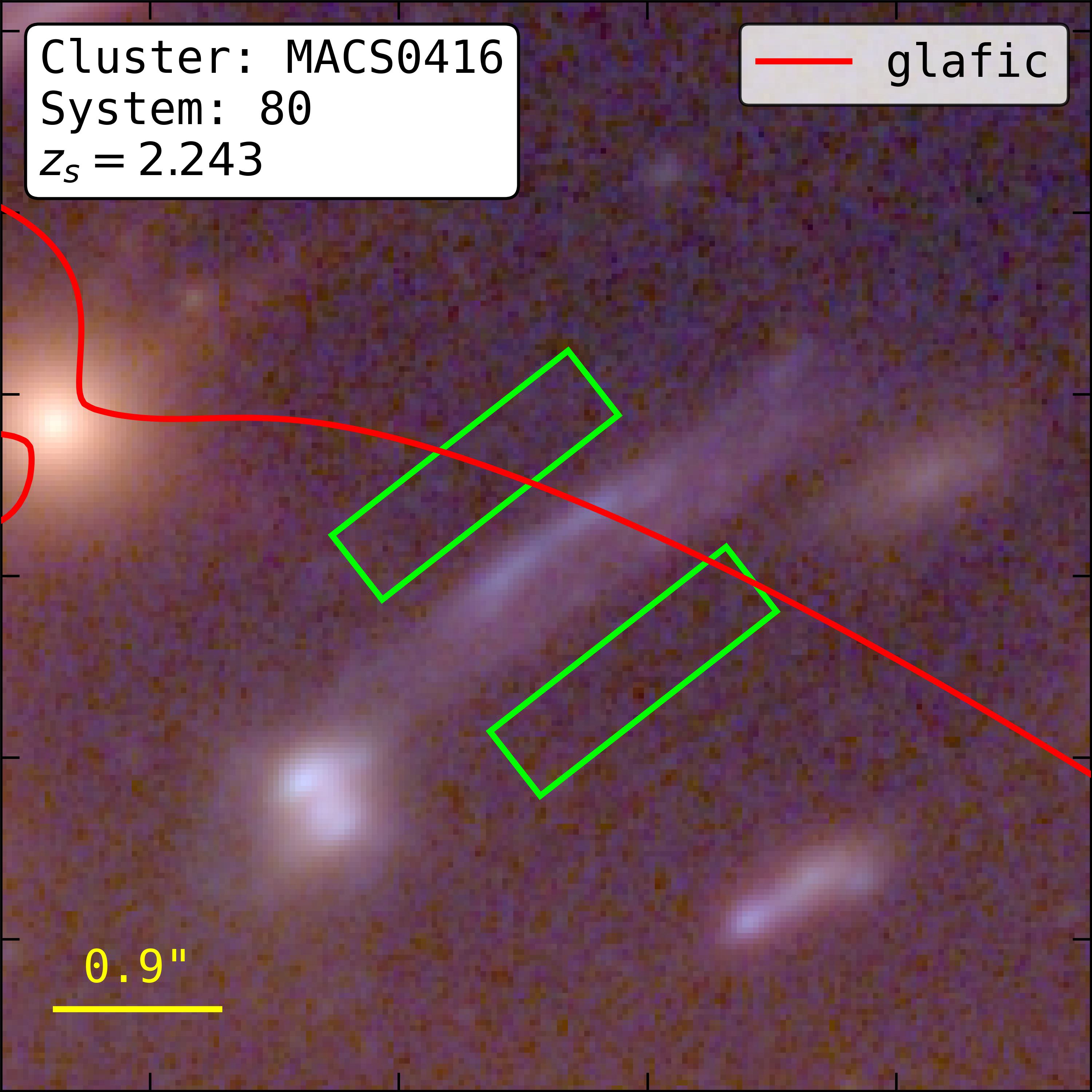}
    \includegraphics[width=4.0cm,height=4.0cm]{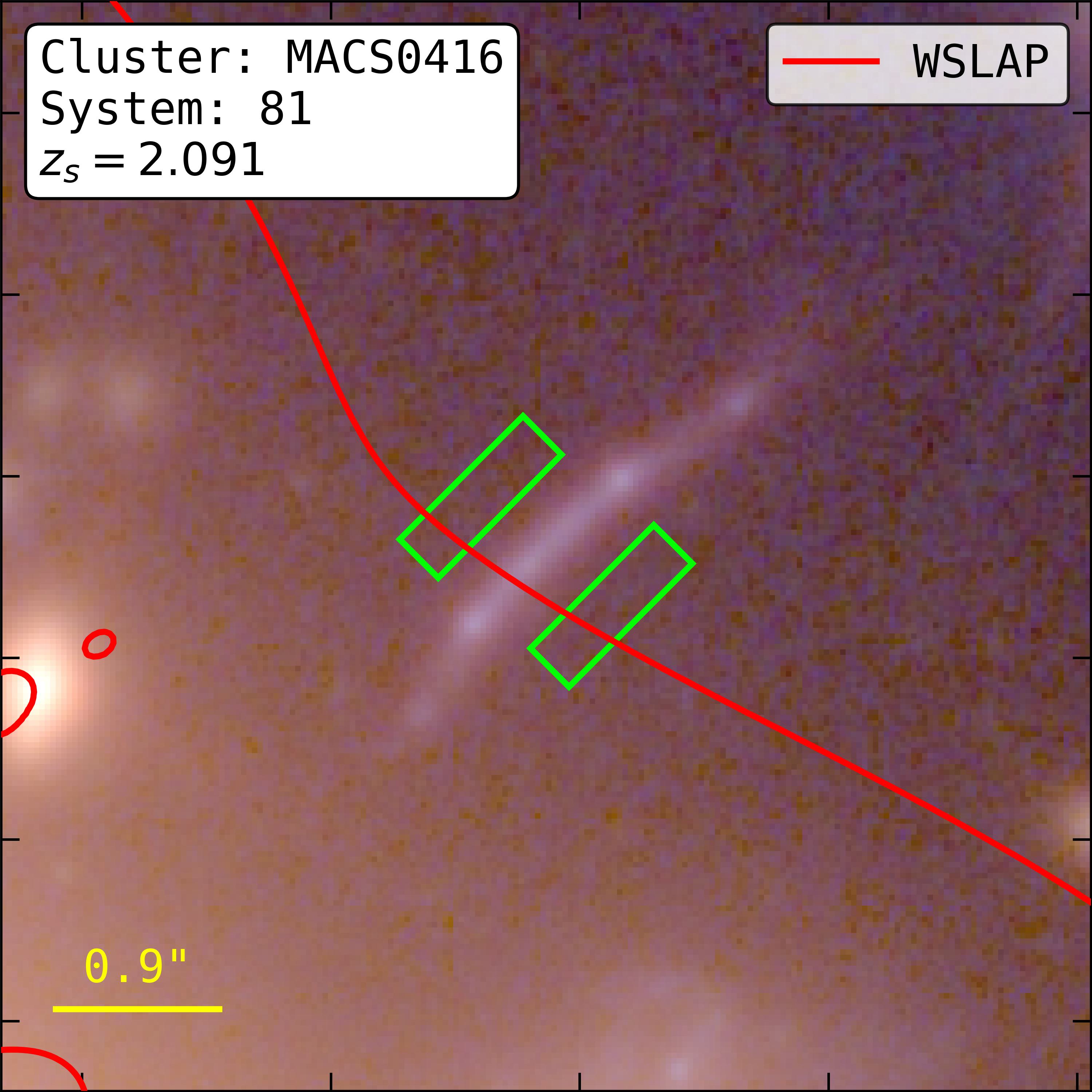}
    \includegraphics[width=4.0cm,height=4.0cm]{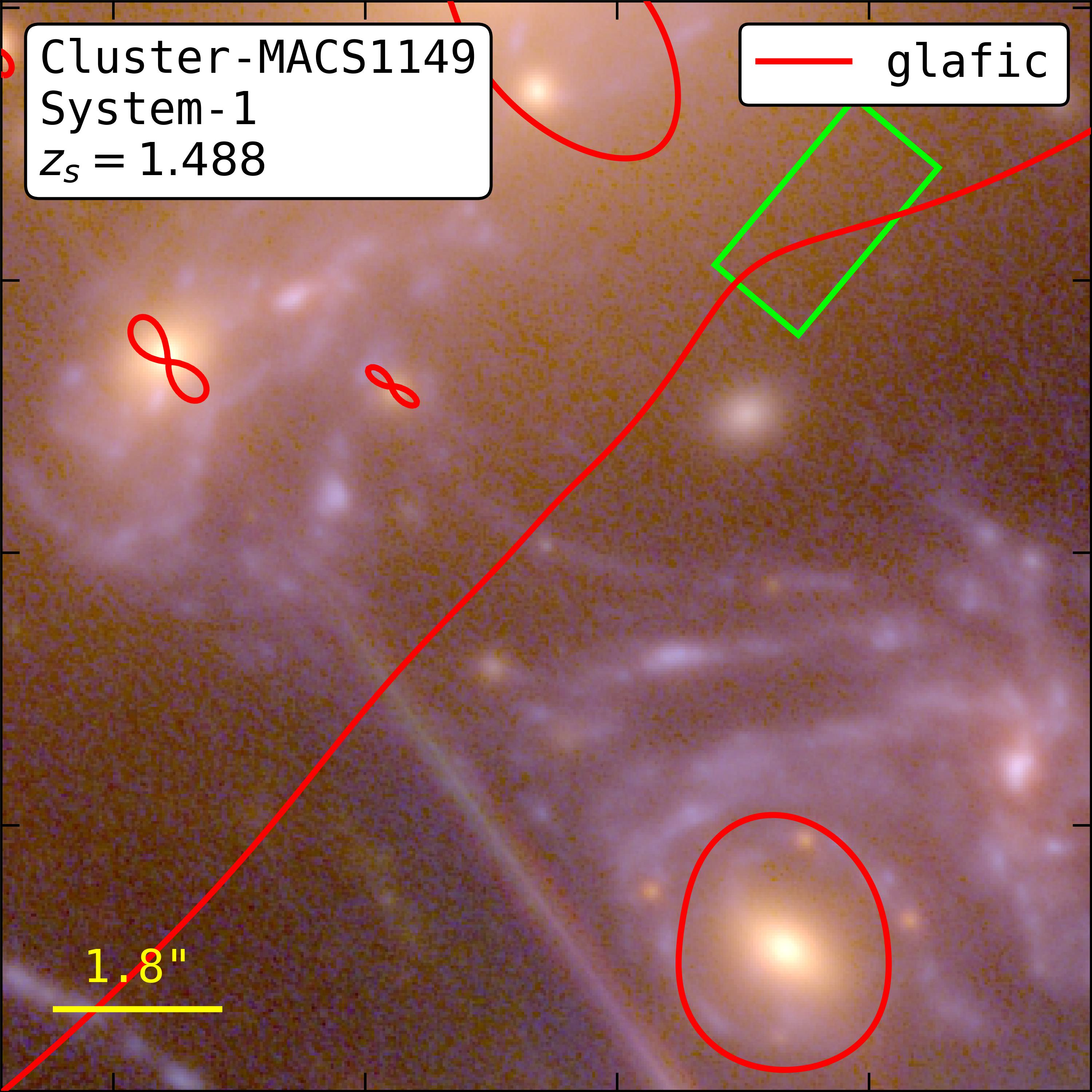}
    \includegraphics[width=4.0cm,height=4.0cm]{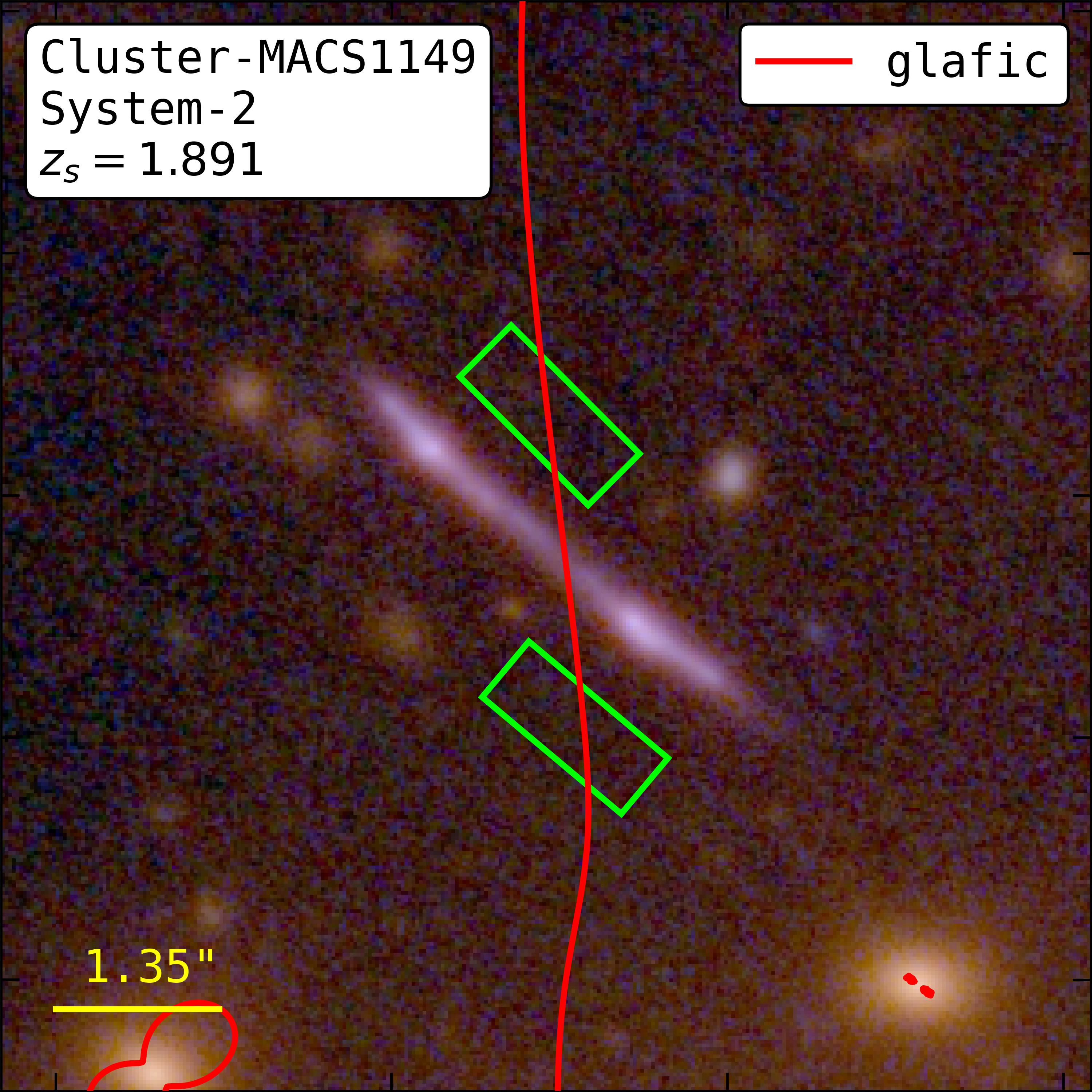}
    \caption{Lensed arcs studied in our current work. In each panel, the name of the galaxy cluster, lensed arc system number (same as shown in Table~\ref{tab:arcs}), and arc redshift are given in the upper-left part. The red curves show the critical curve at the arc redshift corresponding to the lens model (mentioned in the upper-right part of the panel) used in the current work for various estimates for each arc. The green rectangles mark apertures used to estimate the ICL density near each arc. The color images are made using HFF data with~R=F105W+F125W+140W+160W; G=F606W+F814W+F105W; B=F435W+F606W.}
    \label{fig:arcs}
\end{figure*}

\section{Methodology for estimating the CCE rate}
\label{sec:rate}
This section briefly discusses the methodology used to estimate the expected number of CCE rates in Flashlights. 

\subsection{Sample of lensed arcs}
\label{ssec:lensed_arcs}
Lensed arcs used in the current work are listed in Table~\ref{tab:arcs}, and the corresponding color stamps are shown in Figure~\ref{fig:arcs}. We studied a total of seventeen cluster caustic-crossing arcs in Flashlights. The twenty transients detected in the Flashlights program during the two visits are only found in seven of these arcs, whereas the remaining ten did not yield any transients. We refer readers to~\citet{2022arXiv221102670K} for a view of the Flashlights difference imaging showing the transients in the various lensed arcs. 

While this does not affect our analysis or results, we note that this is not a complete list of lensed arcs around cosmic noon in the HFF clusters. For example, there are some other lensed arcs that lie very close to one of the cluster galaxies such that they are heavily contaminated (complicating the analysis and likely also reducing the chances for observing CCEs in them), or pairs of lensed galaxy images that do not necessarily form arcs but can have certain parts with magnification factors~$>10$, which may be sufficient to give rise to observable CCEs from very bright stars. Although their contribution is not expected to be major, a more exhaustive analysis should also consider such systems and is left for future work. 

\subsection{Microlens density}
\label{ssec:stel_den}
Before estimating the stellar population in the source galaxies, one important ingredient for calculating the rate of CCEs is the microlens density in the ICM close to these arcs. We placed multiple rectangular apertures near each lensed arc~(see Figure~\ref{fig:arcs}) to photometrically measure the intracluster light~(ICL) in the seven HFF filters. We used \textsc{bagpipes} to fit a spectral energy distribution~(SED) to the photometry and estimate the ICL surface density near the lensed arcs. \textsc{bagpipes} uses simple stellar population synthesis~(SSPS) models with the Kroupa IMF~\citep{2002MNRAS.336.1188K} along with an exponentially declining star-formation history~(SFH) to do this. The measured stellar microlens density near each lensed arc is shown in column~(5) of Table~\ref{tab:arcs}, and we used the best-fit values for various estimations.

Since all stars in the ICM are essentially stripped from cluster galaxies during hierarchical cluster formation~(except for a very small fraction that might have formed within the ICM), they primarily consist of stars older than a Gyr~\citep[e.g.,][]{2018MNRAS.474..917M}. Hence, typical ICM stellar mass values will be in the mass range~$\sim$[0.08, 1.5]~${\rm M_\odot}$ as all higher-mass stars will have completed their life within roughly 1~Gyr. For simplicity, we assumed that all stellar microlenses have a constant mass of~0.3~${\rm M_\odot}$, similar to the average value of the above mass range (see~\citealt{2004ApJ...613...77S} for the validity of such an approximation). In principle, during stellar stripping, stellar remnants (like white dwarfs, neutron stars, and black holes) will also get stripped from the infalling galaxies and enrich the ICM. These stellar remnants will be constituents of compact dark matter and will contribute to the CCE rate. However, for simplicity, in our current work, we assumed that all dark matter is in the smooth form.

\subsection{Stellar population in the lensed galaxies}
\label{ssec:stel_pop}
For modeling the stellar population in each arc, we fitted an SED to the lensing-corrected photometry (calculated using the HFF observations) of the lensed arcs themselves. We obtained lensing-corrected photometry by magnification correcting each pixel in the lensed arcs and by dividing the total flux by the number of merged images that form the lensed arc. For example, fold arcs have two images, so the division factor will be two. Using \textsc{bagpipes}, we then obtained the star-formation rate (SFR) and other parameters, and fed them into \textsc{spisea} to simulate the unlensed stellar population in each arc and evaluated the number of expected CCEs. We detail below the setup for each of these. 

Since the lensing correction depends on the lens model, we expect to get different values of lensing-corrected flux for different lens models. However, most of the lensed arcs are double-image arcs where the critical curve should pass close to the symmetry point. For each arc, we use the publicly available HFF-v4 lens models that put the critical curve closest to the symmetry point. For most arcs, we found this was best matched by the \textsc{glafic}~\citep{2010PASJ...62.1017O} lens models\footnote{\url{https://archive.stsci.edu/prepds/frontier/}}, which is the one we typically adopt (the explicit model for each arc is designated in Figure \ref{fig:arcs}). For more complicated arcs, it can be very hard to determine the exact position (as well as shape) of the critical curve and the exact image multiplicity in the arc~(for example, the Spock arc, M0416-Arc11, i.e., System~11 in the MACS~0416 galaxy cluster). Also, for such complex arcs, we decided to adopt the \textsc{glafic} lens models in order to maximize consistency with the rest of the sample. The effect of lens models on the number of CCEs is further discussed in Section~\ref{ssec:uncertain_lens}. To mitigate the lens-model-induced error and other systematic effects on the SED fitting, we added an extra 10\% uncertainty to the lensing-corrected photometry in each filter. Here we do not take into account the systematic uncertainty in the macromagnification of each arc coming from variations in the position of the macrocritical curve for different lens models, which can be on the order of another 30--50\% and affects all bands in the same manner~\citep{2015ApJ...801...44Z, 2017MNRAS.465.1030P, 2020MNRAS.494.4771R}. However, for most arcs (especially fold arcs), visual inspection can easily find the extreme outlier lens models, helping us reduce this uncertainty. Below, we briefly mention the \textsc{bagpipes} and \textsc{spisea} setups used to estimate the source stellar population.

\subsubsection{\textsc{bagpipes} setup}
\label{sssec:bagpipes_setup}
We ran \textsc{bagpipes} to fit for the age since star formation started;~($t_{\rm age, max}$), age since star formation stopped~($t_{\rm age, min}$), total stellar mass formed~($M_\star$), metallicity~($Z$), dust~($A_V$), and ionization parameter~($U$) assuming a constant star-formation history~(SFH) and Calzetti dust law~\citep{1994ApJ...429..582C}. We used flat priors for the above quantities with the following ranges: $t_{\rm age, max}\in(0.01, 13)~{\rm Gyr}$, $t_{\rm age, min}\in(0.01, 0.3)~{\rm Gyr}$, $\log[M_\star/{\rm M_\odot}] \in (4, 12)$, $Z/Z_\odot \in (0.01, 1.0)$, $A_{\rm V}\in(0,3)$, and $\log(U)\in(-4,-1)$. Photometry measured from HFF observations is used. Since all of the HFF clusters (except MACS~1149) also have deep MUSE observations~\citep{2021AA...646A..83R}, we also used equivalent line-width measurements (with signal-to-noise ratio S/N $\gtrsim5$) for various lines while running \textsc{bagpipes}. We extracted the SFR~($\psi$) in the last 0.3~Gyr from the full SFH of each lensed arc, as the primary contribution to CCEs in F200LP is expected to come from O- and B-type stars, which only live for short periods~($<0.3$~Gyr). Since our aim is to assess the effect of the IMF on the number of CCEs expected in the different arcs, we ran \textsc{bagpipes} for each arc using two different IMFs, namely Salpeter~($\alpha=2.35$) and top-heavy~($\alpha=1.0$), assuming a stellar mass range of~$[0.1, 100]\:{\rm M_\odot}$. One can, even at the SED fitting stage, look at the resulting fits and try to make a statement about the preferred stellar IMF. However, in this work, we focus on the more direct probe, which is the observed number of CCEs.

\subsubsection{\textsc{spisea} setup}
\label{sssec:spisea_setup}
Once we got the best-fit values for~$(\psi, A_V, Z)$ along with their~$1\sigma$ uncertainties from \textsc{bagpipes}, we fed these values into \textsc{spisea} to generate the rest-frame stellar population that was formed in the last 0.3~Gyr. To be specific, we generated 10~realizations for each arc sampling the $1\sigma$ ranges on the above parameters; this would then reflect the range of uncertainties in our CCE estimate from these parameters. The \textsc{bagpipes} runs are done with the same IMF that was adopted in the \textsc{spisea} setup (i.e., Salpeter or top-heavy). To generate isochrones and evolve the stellar population, we used \textsc{mesa} Isochrones \& Stellar Tracks~\citep[\textsc{mist-v1};][]{2016ApJ...823..102C}, \textsc{get$\_$merged$\_$atmosphere} as the atmosphere model, and \textsc{RedLawHosek18b} as the extinction law, all of that are available in \textsc{spisea}. In addition, for simplicity, we switched off stellar multiplicity. Since \textsc{bagpipes} gives~$A_V$ in the output, we use a ~$A_K = 0.062\,A_V$ scaling following~\citet{2008ApJ...680.1174N} to input dust-extinction in \textsc{spisea}. With the above assumptions, we divided the last 0.3~Gyr history into 1~Myr bins and evolved the stellar population generated in each bin accordingly. At the end of this process, we obtained the apparent magnitude~($m_{\rm AB}$) in the F200LP filter for each star in the simulated stellar population from \textsc{spisea}.

\subsection{From stellar population to the expected number of CCEs}
\label{ssec:star_to_cce}
Assuming that the simulated stellar population is randomly distributed in the lensing-corrected area of the arc, the \emph{observed} number of CCEs in a given arc at any time is given as, 
\begin{equation}
\begin{split}
    N_{\rm CCE,arc} = & \int_{m_{\rm AB}} \int_{\mu_{\rm m}}  \: d m_{\rm AB} \: d\mu_{\rm m} \: n(m_{\rm AB}, \mu_{\rm m},\alpha) \quad \times \\ 
              & \frac{dA_s (\mu_{\rm m})}{d\mu_{\rm m}} \Delta t (m_{\rm AB}, \mu_{\rm m}) \: R_{\rm CCE} (\mu_{\rm m}, n_{\rm micro}, v, \theta),
    \label{eq:n_cce}
\end{split}
\end{equation}
where~$n(m_{\rm AB}, \mu_{\rm m}, \alpha)$ is the number density of source stars with a magnitude~$m_{\rm AB}$ such that they can be observed with a macromagnification of~$\mu_{\rm m}$, and~$dA_s / d\mu_{\rm m}$~($=(1/\mu_{\rm m}) dA_i / d\mu_{\rm m}$) represents the lensing-corrected source area with macromagnification~($\mu_{\rm m}$) and~$dA_i$ is the corresponding image-plane area. $\Delta t$ represents the observer-frame time when the background star will remain above the detection threshold, and $R_{\rm CCE}$ is the average CCE rate (per unit observer time) in this area. The integral on~$\mu_{\rm m}$ runs from the minimum to the maximum value of macromagnification in the lensed arc. Similarly, the integral on~$m_{\rm AB}$ runs over the apparent magnitudes of the whole stellar population. The CCE rate, in addition to macromagnification~($\mu_{\rm m}$), also depends on the number of microlenses~($n_{\rm micro}$, and their density) in the image plane, relative velocity~($v$) of the source star with respect to the observer, and its direction of motion~($\theta$) with respect to the macrocaustic. 

The presence of microlenses in the ICM introduces microcritical curves and microcaustics in the image and source planes, respectively. When a background star crosses one of these microcaustics, a pair of microimages (on one of the microcritical curves) appears or disappears, having large magnification factors. While crossing the macrocaustic, the total magnification of all of the microimages of the background star is referred to as the peak magnification. The exact value of peak magnification depends on the local properties of the macrolens and the microlenses~\citep[e.g.,][]{2017ApJ...850...49V, 2018ApJ...857...25D, 2018PhRvD..97b3518O, 2019A&A...625A..84D}, the radius of the star, and the number of microcaustics that the star is crossing at any time (which is usually one at a time). Very close to the macrocritical curve, these microcritical curves overlap with each other, forming a corrugated network~\citep{2017ApJ...850...49V}. The CCE peak magnification saturates inside the corrugated network while the network's width depends on the microlens surface density and local rate of change of macromagnification near the macrocritical curve. The magnification~($\mu_{th}$) at the boundary of the corrugated network is given by
\begin{equation}
 \mu_{th} = \frac{\Sigma_{\rm crit}}{\Sigma_*},
\end{equation}
where~$\Sigma_{\rm crit}$ and $\Sigma_*$ are the critical surface density and microlens density, respectively. The above condition comes from the demand that at the edge of the corrugated network, the microlensing optical depth will be equal to 1. An interesting aspect of the above equation is that it does not depend on the microlens mass function~(a result of the fact that optical depth is also independent of the microlens mass function and only depends on the average mass) and only depends on the total microlens surface density~\citep[e.g.,][]{1995MNRAS.276..103L,Wyithe2000}. Note that in the presence of microlenses, in principle, one would need to simulate explicit light curves to determine the magnification factor for each peak. However, in our current work, for simplicity, we assume that the peak magnification is given by Equation~(26) of \citet{2018PhRvD..97b3518O},
\begin{equation}
    \mu_{\rm peak} = \mu_{\rm t} \mu_{\rm r} \left( \frac{\theta_{\rm E,*}}{\sqrt{\mu_{\rm t}}\beta_{R}}\right)^{1/2},
    \label{eq:mu_peak}
\end{equation}
where~$\theta_{\rm E,*}$ is the Einstein angle corresponding to the microlens, $\beta_R$ is the angular radius of the star, and~$\mu_t (\mu_r)$ represents the tangential~(radial) component of macromagnification. 

As mentioned above, the peak magnification saturates inside the corrugated network and is not well described by the above formula. For simplicity, however, we used the corresponding peak magnification value as the peak magnification also inside the corrugated network. For some arcs, the inferred peak magnification can be extremely high,~$(\gtrsim10^4)$, which is not the case for a typical CCE inside the network (since the above formula is derived for low optical depths). There, we limited the peak magnification to a conservative value of~$5\times10^3$. From Equation~\eqref{eq:mu_peak}, we can see that the peak magnification also depends on the size of the background source star, and, for simplicity, we chose a radius of 20~${\rm R_\odot}$ in current work since we primarily expect CCEs from O- and B-type stars in F200LP from around cosmic noon and larger stars like red supergiants will be brighter at longer wavelengths (such as in JWST images; \citealt{2023A&A...672A...3D}). 

To determine the average CCE rate, we need to calculate the average time in the observer frame between two consecutive CCEs per background star, which will depend on the microlensing optical depth and the macromagnification factor. As we go from the image plane to the source plane, the overall area is squeezed in different directions based on the magnification factor~($\mu_m = \mu_t \times \mu_r$; e.g.,~\citealt{1992grle.book.....S}), increasing the microcaustics density (by the same magnification factors) compared to microcritical curves. Assuming $\mu_t \gg \mu_r$ and that the elongation of lensed arcs is perpendicular to the network microcritical curves, a source star moving perpendicular to the macrocaustic~(i.e., the direction along which the observed arc is elongated) would cross the maximum number of microcaustics, and a minimum number if moving parallel to the macrocaustic direction. The opposite of this will happen for~$\Delta t$;  CCEs will be short-lived if the source is moving perpendicular and will last longer if the source is moving parallel to the macrocaustic direction. The number of CCEs that the source star will cross also depends on the size of the microcaustic in these directions because it determines the distance up to which microlenses can lie such that the corresponding microcaustics would be in the path of the source star. The size of microcaustics for a microlens on the positive- and negative-parity sides of the macrocaustic is given as~\citep{2018PhRvD..97b3518O}
\begin{equation}
\begin{split}
    \beta_{\rm minima, x} &\approx \frac{1}{\sqrt{\mu_r}} \theta_{\rm E,*}, \qquad
    \beta_{\rm minima, y} \approx \frac{\sqrt{\mu_t}}{\mu_r} \theta_{\rm E,*}  \\
    \beta_{\rm saddle, x} &\approx \frac{1}{\sqrt{\mu_r}} \theta_{\rm E,*},  \qquad
    \beta_{\rm saddle, y} \approx \frac{1}{\mu_r} \sqrt{\frac{\mu_t}{8}} \theta_{\rm E,*},
    \label{eq:c_widths}
\end{split}
\end{equation}
where~$\theta_{\rm E,*}$ is the Einstein angle corresponding to the microlens, and $\beta_{\rm minima, x/y}$ and $\beta_{\rm saddle, x/y}$ represent the sizes of microcaustics parallel/perpendicular to the macrocaustic direction corresponding to a microlens on the minima and saddle side of the microcritical curve, respectively.

Once we determine the CCE rate in the two directions~(i.e., source moving perpendicular and parallel to the macrocaustic), we can determine the CCE rate in any other direction using~\citep{1993A&A...268..501W}
 \begin{equation}
    R_{\rm CCE}(\phi) = \sqrt{\frac{a^2 (1-e^2)}{1 - e^2 \cos^2(\phi)}},
\end{equation}
where~$\epsilon=\sqrt{1-b/a}$ with~$a$ and~$b$ representing the CCE rate for a source star moving perpendicular~($\phi=0$) and parallel~($\phi=\pi/2$) to microcaustic, respectively. The applicability of an ellipse equation to determine the average CCE rate in any arbitrary direction can be understood from the fact that a circular region (of unit radius) in the image plane will map back to an ellipse in the source plane with its major and minor axes depending on the~$\mu_t$ and $\mu_r$ values. Hence, uniformly distributed microlenses in the image plane will lead to uniformly distributed microcaustics in the source plane but with respect to the elliptical geometry, implying that the average CCE rate will also follow the same elliptical geometry. 

One additional fact that we also take into account while estimating the rate of CCEs is that a microlens on the positive-parity side of the critical curve leads to a diamond-shaped microcaustic in the source plane, whereas a microlens on the negative-parity side leads to two triangular-shaped caustics in the source plane~\citep[e.g.,][]{1984A&A...132..168C}. Hence, a source star will lead to a factor of two more CCEs on the negative-parity side compared to the positive-parity side when it is moving perpendicular to the macrocaustic. Hence, the CCE rate corresponding to a source star will be
\begin{equation}
    R_{\rm CCE}(\mu_{\rm m}) = 
    \begin{cases}
    2\:R_{\rm minima} + 4\:R_{\rm saddle}, & {\rm if}~\phi=0    \\
    2\:R_{\rm minima} + 2\:R_{\rm saddle}, & {\rm if}~\phi=\pi/2,
    \end{cases}
    \label{eq:r_cce}
\end{equation}
where $R_{\rm minima}$ is the number of diamond-shaped microcaustics and~$R_{\rm saddle}$ is the number of triangular-shaped caustics that a source star crosses per unit time corresponding to microlenses on the positive- and negative-parity sides of the macrocritical curve. 

\subsection{Comparison with ray-tracing simulations}
\label{ssec:comp}
To validate our formalism, we compared the peak magnification~($\mu_{\rm peak}$; Equation~\eqref{eq:mu_peak}) and CCE rate~($R_{\rm CCE}$) estimated using our approximated method~(Equation~\eqref{eq:c_widths} -- \ref{eq:r_cce}), with explicit ray-tracing simulations using our ray-tracing code presented by \citet{2022MNRAS.514.2545M}. While ideally, the true peak magnification and CCE rate could be derived for each arc directly from ray-tracing simulations, this cannot be done for each arc in the sample because it is a very time-consuming task, which would become an even bigger challenge when progressively more arcs are incorporated in larger samples in the future. Instead, we only estimated here the typical bias of our simplified (and more time-efficient) method compared to ray-tracing simulations, and can then correct for this factor. Here we run ray-tracing simulations for a few fiducial arcs with typical values and with fifty realizations per configuration.  

\begin{figure}
    \centering
    \includegraphics[width=8.5cm,height=10cm]{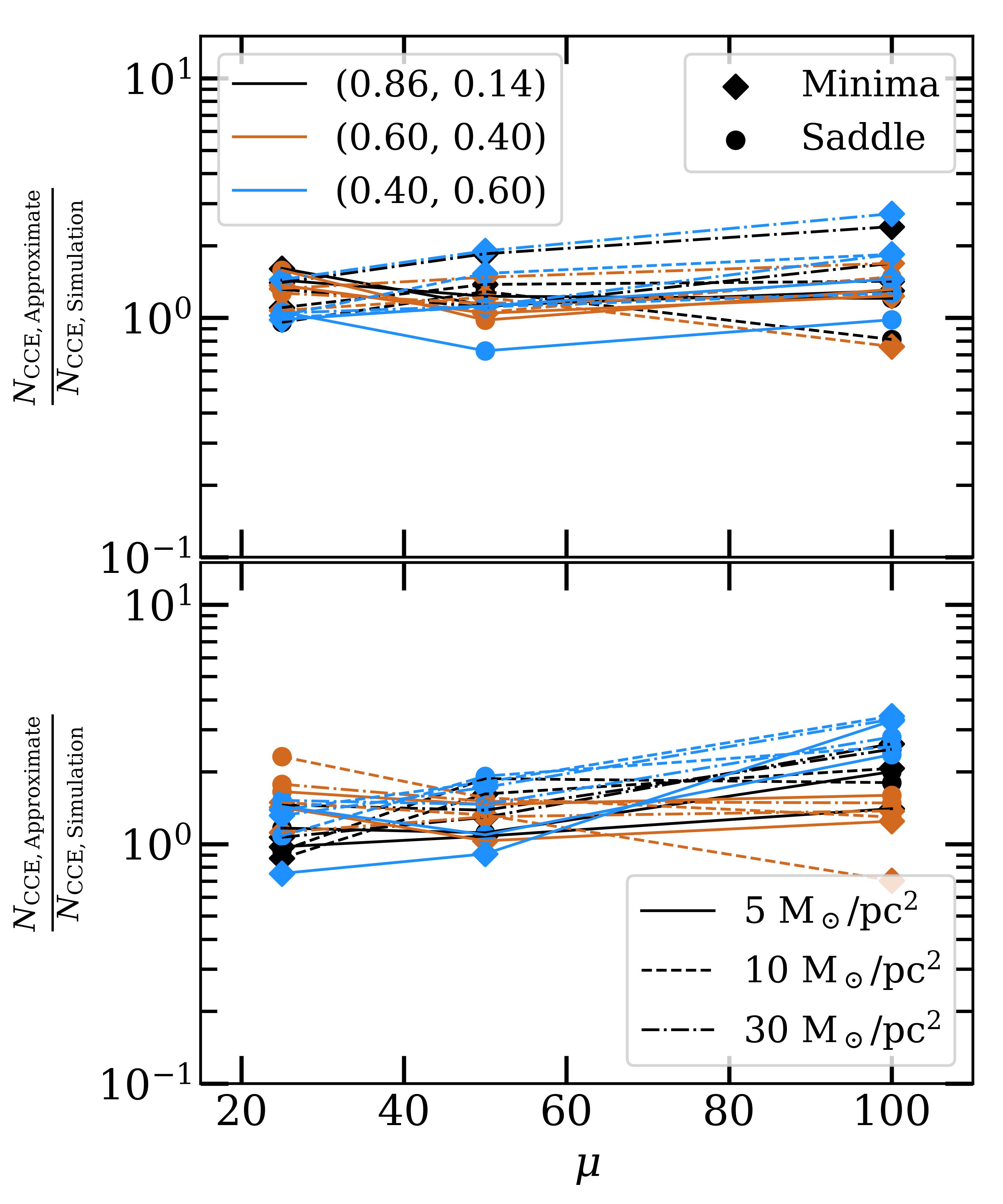}
    \caption{Validation of our simplified method for calculating the rate of CCEs. The abscissa and ordinate axes represent the macromagnification in the image plane and the ratio of the approximate number of CCEs estimated using our method to the number of CCEs from explicit ray-tracing simulations, respectively. We show a comparison for three stellar density values. The top (bottom) panel indicates this ratio for a source moving along the~$\phi=0$~($\phi=\pi/2$) path. As can be seen, the number of predicted CCEs from our simplified method overall agrees well with the estimates from the explicit simulation, overestimating the number of events by a factor of 1--3 -- mainly due to the fact that in reality, some microcaustics would merge into larger caustics, which is not taken into account in our calculation.}
    \label{fig:compare}
\end{figure}

The results corresponding to a CCE rate comparison from our approximated method to simulations are shown in Figure~\ref{fig:compare}. The top (bottom) panel represents the fraction of the CCE rate from the above method compared to simulations for a source moving along the~$\phi=0$~($\phi=\pi/2$) path. We chose three different values of image plane macromagnification~(i.e.,~$\mu_{\rm m}=25, 50, 100$) and for each~$\mu_{\rm m}$ value we use three microlens density values~(i.e.,~$\Sigma_* = 5, 10, 30~{\rm M_\odot}/{\rm pc^2}$) assuming a microlens mass of~$0.3~{\rm M_\odot}$. Similarly, to determine the dependency of macroconvergence~($\kappa_{\rm m}$) and shear~($\gamma_{\rm m}$) values, we again choose three different sets of~$(\kappa_{\rm m}, \gamma_{\rm m})$ values and vary macroconvergence to reach the desired macromagnification value\footnote{Here for simplicity, we only varied the macroconvergence~($\kappa_{\rm m}$), but close to actual macrocritical curves, macroshear~($\gamma_{\rm m}$) will also vary.}. We can see that our approximated method gives the number of CCEs within a factor of about 1--3 compared with simulations, which is likely a result of not accounting in our simplified calculation for the fact that some microcaustics will merge together, especially at high magnifications. The differences are higher microlens density and magnification, i.e., for~$30{\rm M_\odot/pc^2}$ at~$\mu=100$. However, since maximum microlens density is~$33{\rm M_\odot/pc^2}$ for our sample of arcs, we do not expect this to lead to significant bias in our results.

\begin{figure}
    \centering
    \includegraphics[width=8.5cm,height=6.5cm]{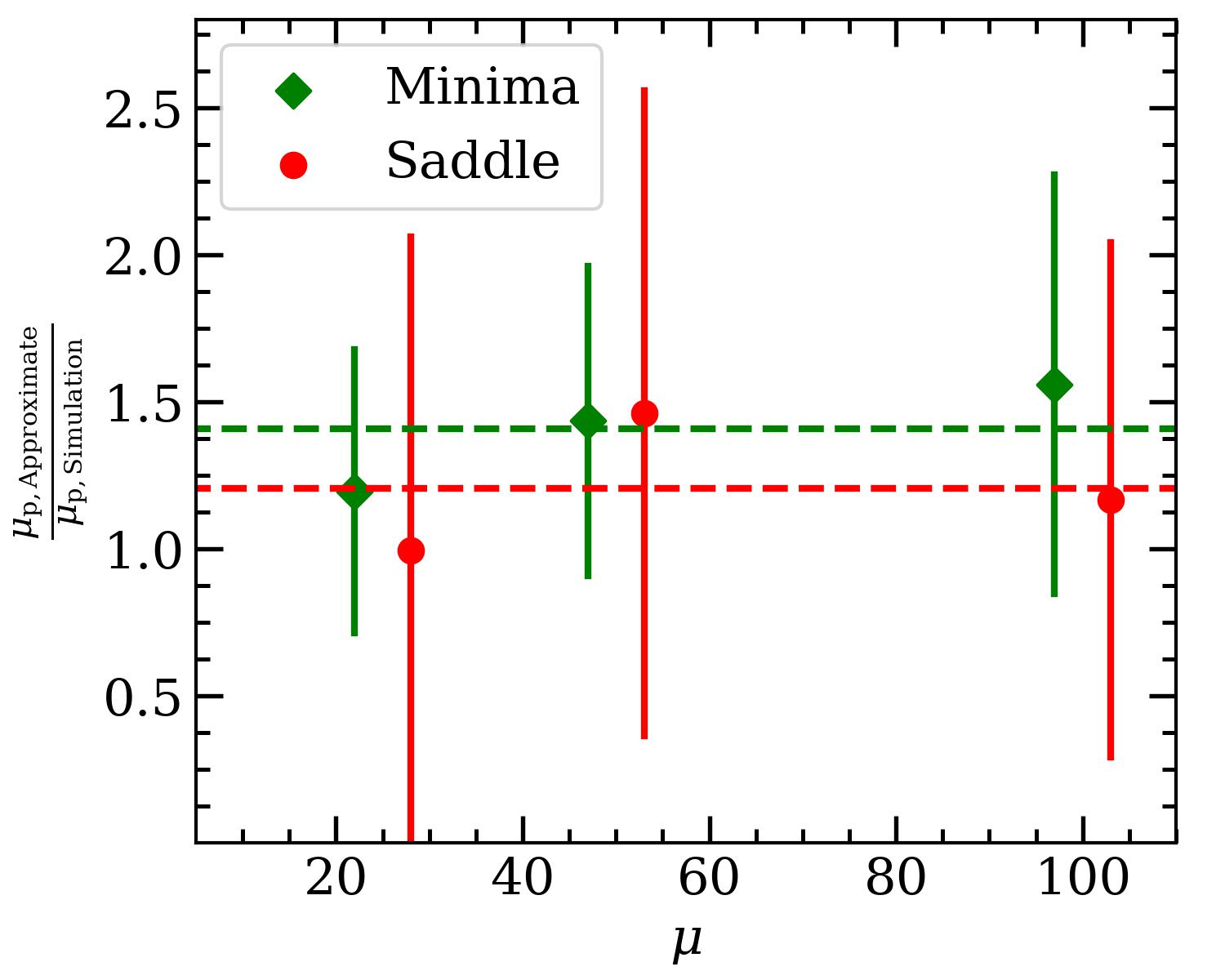}
    \caption{Comparison between peak magnification. The abscissa and ordinate axes represent the macromagnification in the image plane and the ratio of peak magnifications estimated using Equation~\eqref{eq:mu_peak} and average peak magnification from simulations, respectively. The solid green diamonds and red circles correspond to the minima and saddle sides of the macrocritical curve, respectively. The error bars represent the 1$\sigma$ variation in peak magnification around the average value in simulations. The horizontal dashed curves represent the average of the solid points in the same color.}
    \label{fig:mu_peak}
\end{figure}

The results corresponding to peak magnification comparison are shown in Figure~\ref{fig:mu_peak}. Here we only consider one value of~$(\kappa, \gamma)=(0.86, 0.14)$ and~$\Sigma_* = 30~{\rm M_\odot}/{\rm pc^2}$. We see that Equation~\eqref{eq:mu_peak} slightly overestimates the average~$\mu_{\rm peak}$ value on both the minima and saddle side, and the overestimation reduces at lower~$\mu_{\rm m}$ values. At the same time, as expected, we also notice that the variation in~$\mu_{\rm peak}$ values is larger on the saddle side. The horizontal dashed curves represent the average of the solid points in the same color. From the plot, we can infer that Equation~\eqref{eq:mu_peak} overestimates the average~$\mu_{\rm peak}$ values with a factor of~$\lesssim1.5$. 

\section{Results}
\label{sec:results}

\subsection{Constraints on the IMF from the number of CCEs}
\label{ssec:ssps_bound}
\begin{figure}
    \centering
    \includegraphics[width=8.0cm,height=6.0cm]{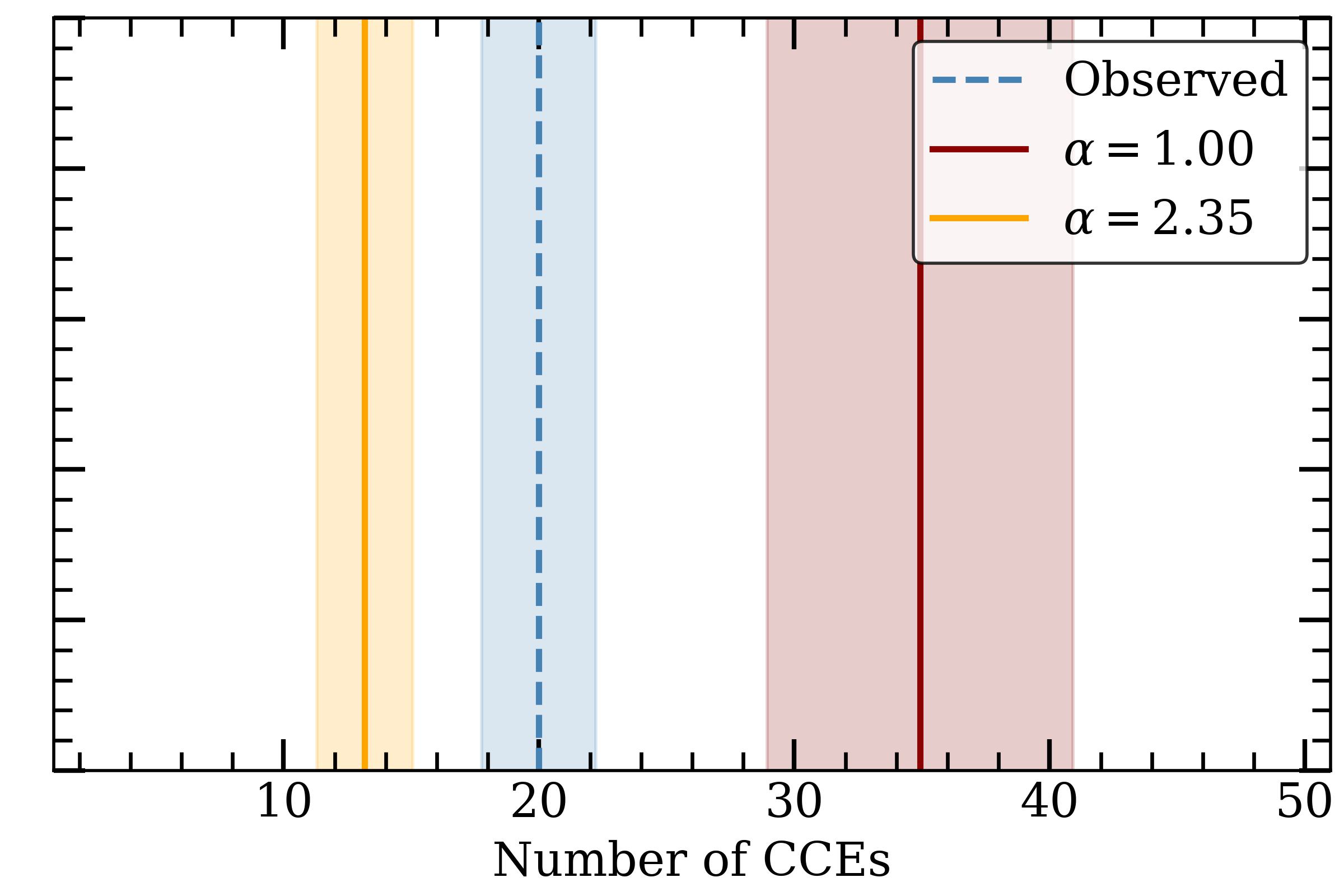}
    \caption{Predicted number of CCEs in the Flashlights survey for two different IMFs used in the current work. The vertical blue dashed line represents the observed number of transients in Flashlights, and the shaded region around it marks the~$1\sigma$ Poisson error. The yellow and red vertical lines represent the expected number of CCEs in the Flashlights survey for Salpeter~($\alpha=2.35$) and top-heavy~($\alpha=1.0$) IMFs, respectively, and the shaded region around these lines marks the~$1\sigma$ uncertainties.}
    \label{fig:estimated_cce}
\end{figure}

We used the formalism described above to obtain the expected CCE rate in each arc and the total number of CCEs expected in the Flashlights survey, and compared it with the observed number. To this end, we looped over the number of stars generated in the last 0.3~Gyr. To get macromagnification~$(\mu_{\rm m})$ values for each star, we used the fact that in the source plane, the differential magnification probability distribution function~(PDF) goes as~$1/\mu_{\rm m}^3$ (i.e., ~$dP/d\mu_{\rm m} \propto 1/\mu_{\rm m}^3$) and generated a random~$\mu_{\rm m}$ value following this PDF. Once we have~$\mu_{\rm m}$, we determined the corresponding peak magnification~($\mu_{\rm peak}$). For assigning a final peak magnification to a source star, we drew a (uniform) random value from a range of~$\mu_{\rm peak}\pm\mu_{\rm peak, std}$, where~$\mu_{\rm peak, std}$ is the average of all $1\sigma$ ranges shown in Figure~\ref{fig:mu_peak}. We also assigned a random direction~($\phi$) to each star, as it determines the period~($\Delta t$) when the star will remain observable. Based on this, the average expected number of CCEs at any time for each arc is mentioned in Table~\ref{tab:arcs}. Since Flashlights observed all of these galaxy clusters two times, the total number of expected CCE during the Flashlights observations for a given arc will be twice what is listed in Table~\ref{tab:arcs}. For Flashlights, with a Salpeter~($\alpha=2.35$) IMF, the total number of CCEs is~$13.18\pm1.87$, whereas a top-heavy~($\alpha=1.0$) IMF gives~$34.93\pm5.99$ CCEs with the uncertainties covering the $1\sigma$ range. From Figure~\ref{fig:estimated_cce}, it seems that, on average, a Salpeter IMF underpredicts the number of CCEs by about a factor of~$\sim0.7$, whereas the top-heavy IMF overpredicts by a factor of~$\sim1.75$, assuming that all transients detected in Flashlights correspond to lensed stars. This implies that the average IMF around cosmic noon is steeper than~$\alpha=1.0$ but shallower than the well-known Salpeter IMF. These results, however, are prone to various sources of uncertainties in Section~\ref{sec:uncertain}. 
It is important to note that while we focus on arcs around cosmic noon, nearly all of the observed and predicted CCEs belong to lensed arcs at~$z\lesssim1$. This implies that constraints based on the observed CCEs will also be mostly applicable to similar source redshifts. On the other hand, the non-detection of CCEs at~$z\gtrsim2$ in Flashlights observations, in line with our estimates, also supplies constraints on the IMF slope and implies that~$\alpha\in[1, 2.35]$ is an allowed range. Slope values outside this range, however, cannot be ruled out.

Focusing on individual arcs (see Table~\ref{tab:arcs}), the expected number of CCEs in Flashlights observations of the Dragon arc~(A370-Arc2) is~$\sim8$--9 with the Salpeter IMF, which agrees well with observations, whereas a top-heavy IMF produces twice as many CCEs as Salpeter. On the other hand, for the Warhol arc~(M0416-Arc28), our method with the Salpeter IMF predicts~$\sim0.3$--0.4 CCEs, whereas observations led to~four events\footnote{Out of these four events, two have S/N $<3$~\citep{2022arXiv221102670K}, whereas we only calculate the number of CCEs with S/N~$>3$.}. With a top-heavy IMF, we get the expected number of CCEs to be~3--4, agreeing well with observations. However, it is important to note that the slope estimates are subject to various uncertainties. For example, we used a microlens density of~$30~{\rm M_\odot/pc^2}$ for the Warhol arc, whereas~\citet{Palencia2025InPrep} uses a microlens density of~$60~{\rm M_\odot/pc^2}$. If we use the same microlens density, we see an increase of a factor of two in the number of expected CCEs. Another example where our method seems to underestimate the observed number of CCEs is the Spock arc~(M0416-Arc11). The Spock arc led to two CCEs in Flashlights observations. However, our method gave zero even if we assume a top-heavy IMF. One reason could be the fact that the exact location of the macrocritical curve is undetermined, and in fact, there may be several crossings across the arc, as some models suggest. A more detailed analysis will be warranted for this arc.

\begin{figure*}[!ht]
    \centering
    \includegraphics[width=5.3cm,height=3.4cm]{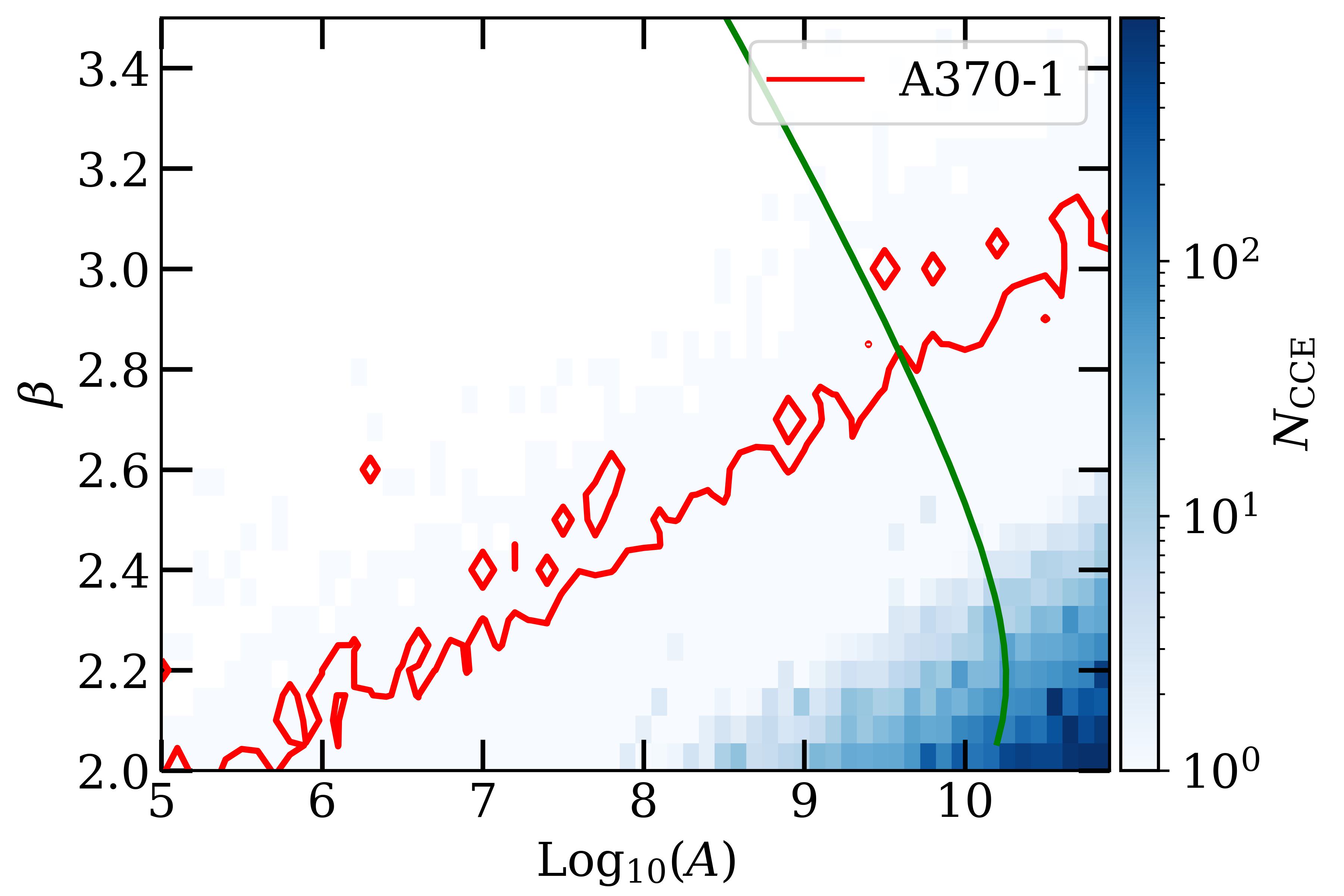}
    \includegraphics[width=5.3cm,height=3.4cm]{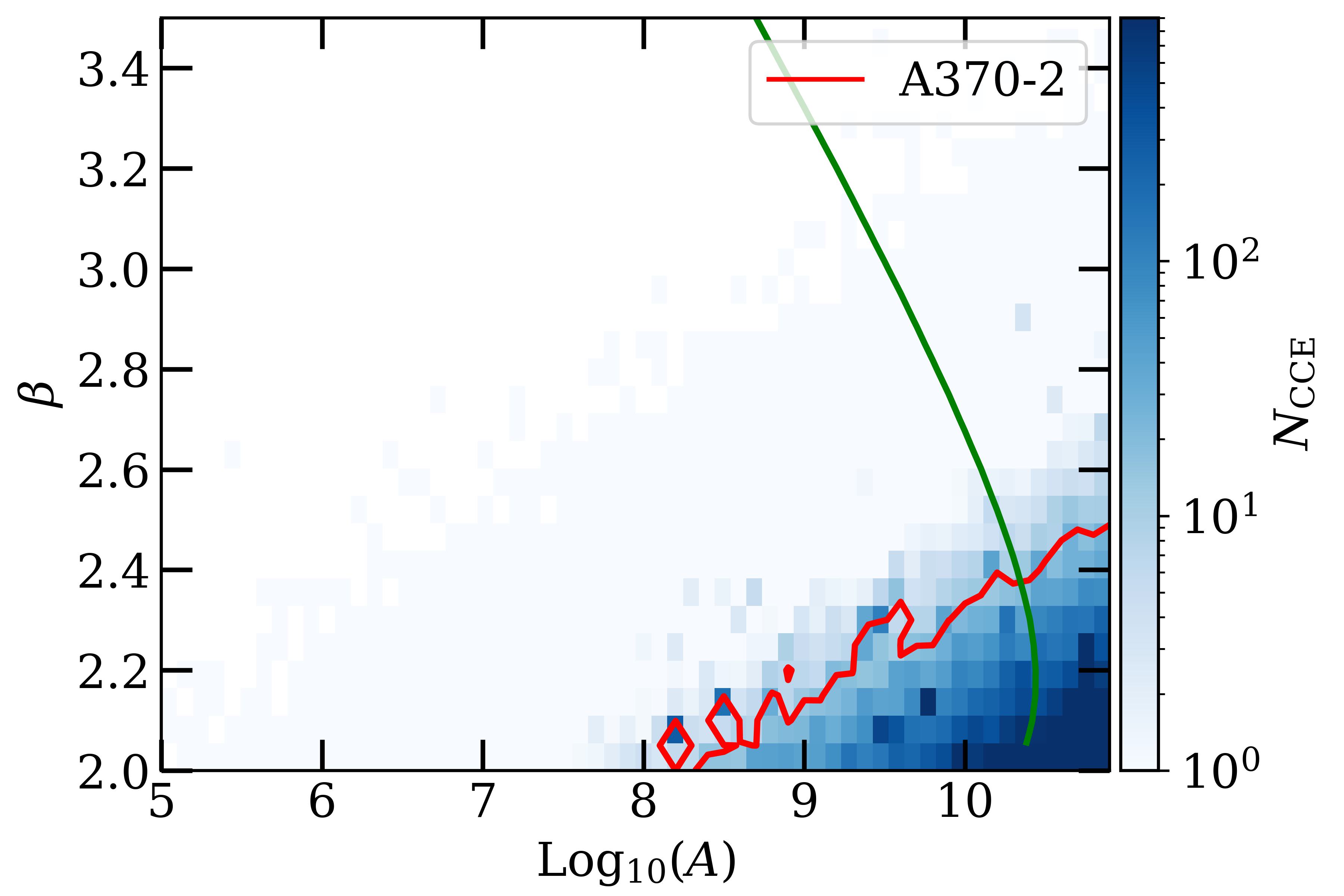}
    \includegraphics[width=5.3cm,height=3.4cm]{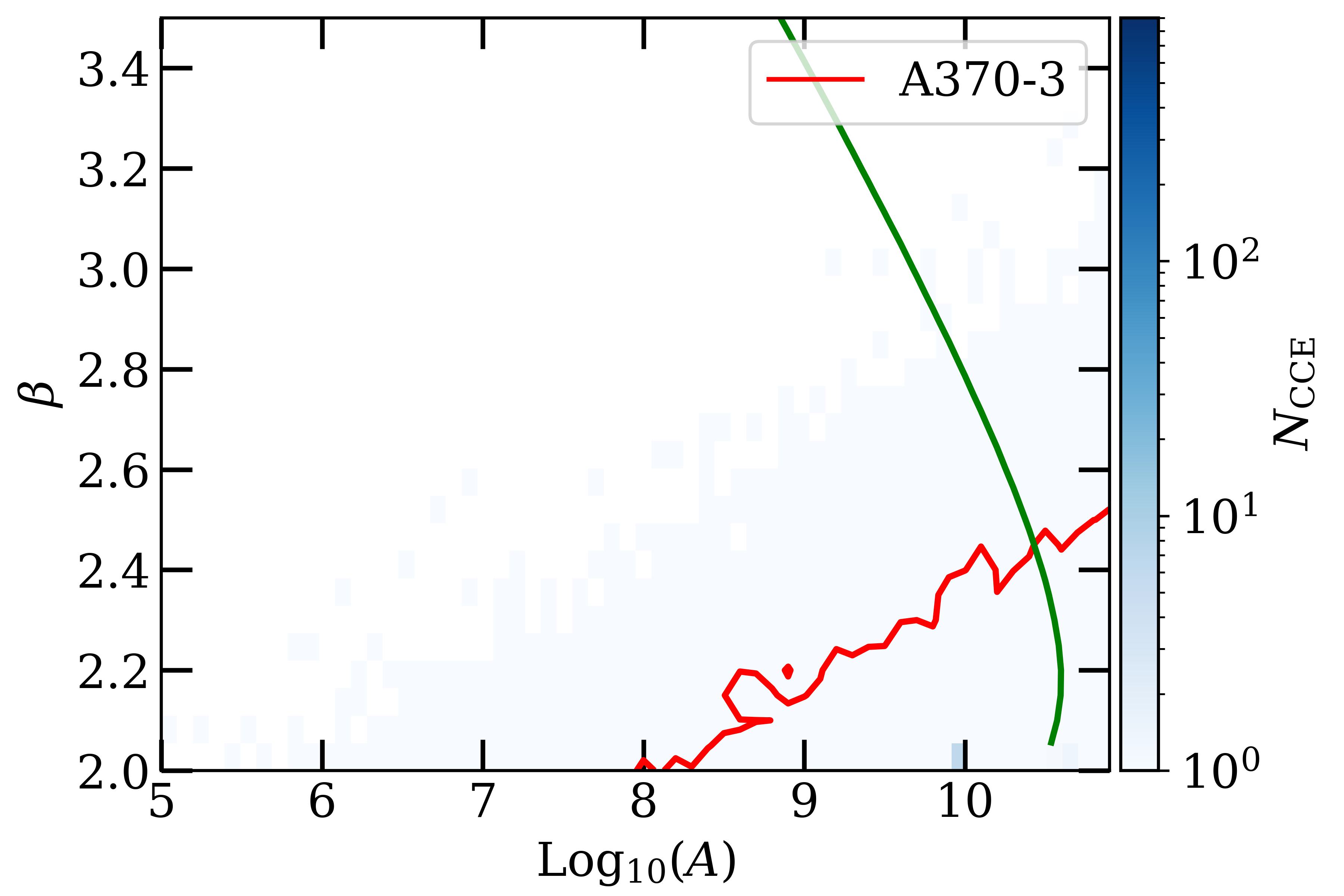}
    \includegraphics[width=5.3cm,height=3.4cm]{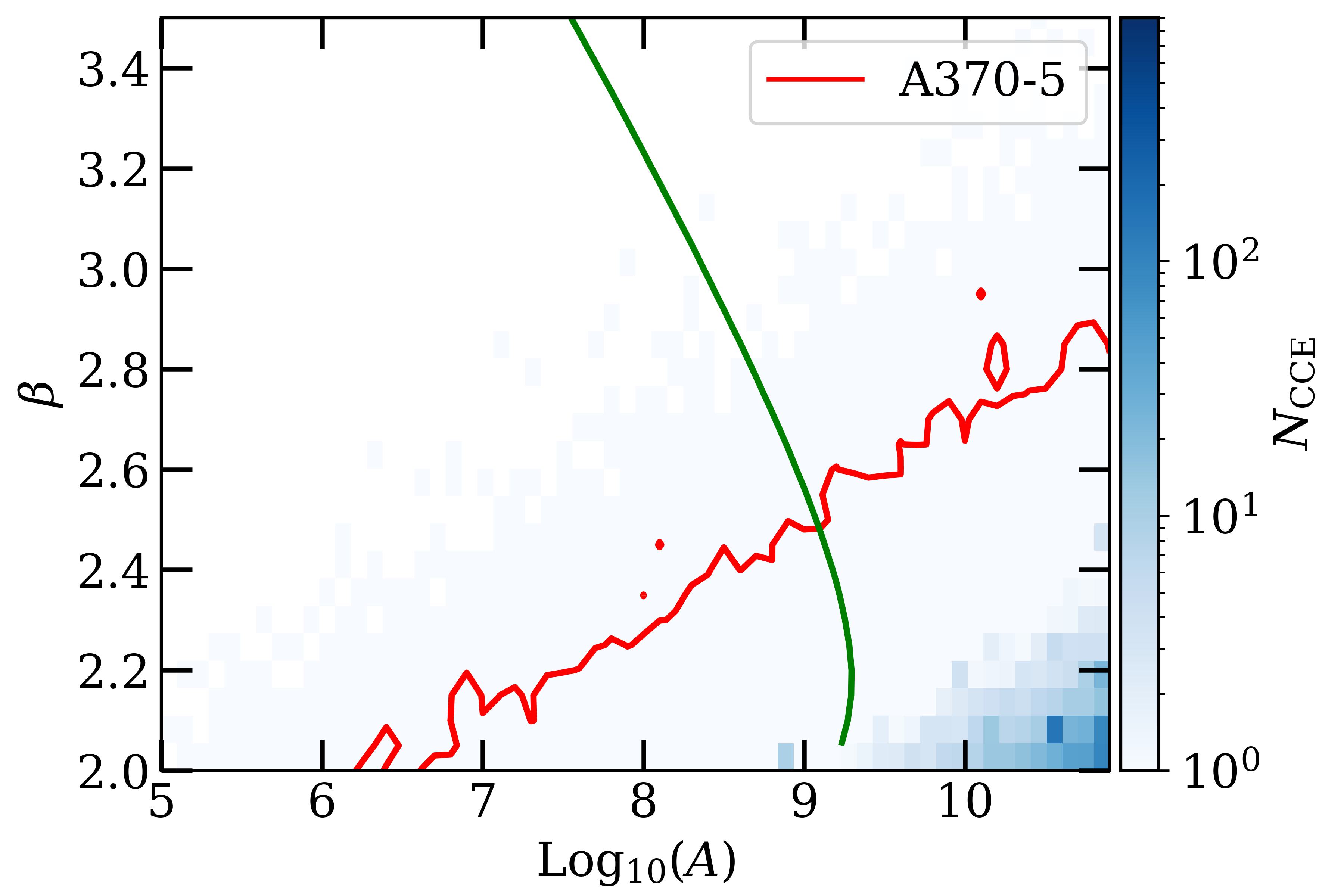}
    \includegraphics[width=5.3cm,height=3.4cm]{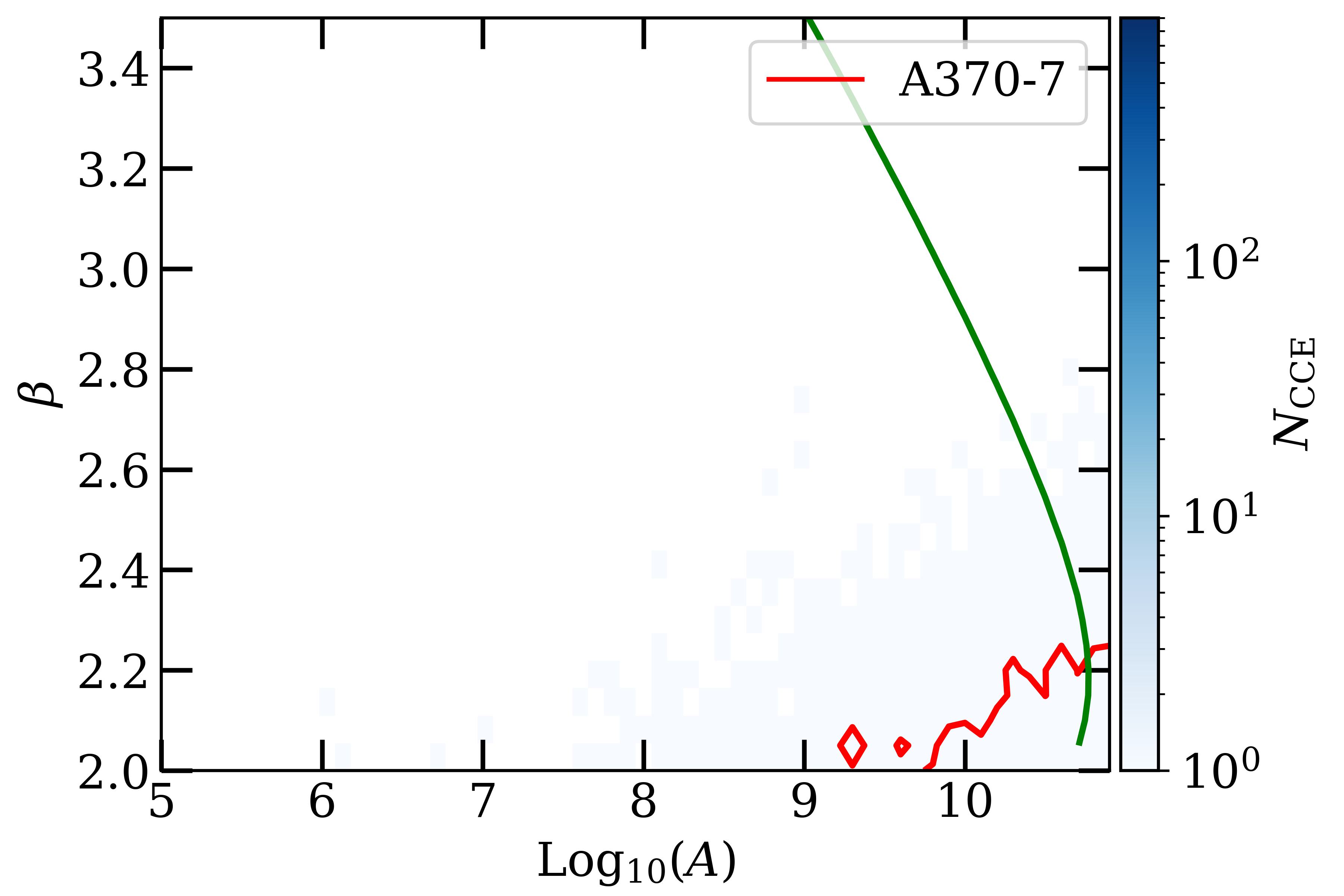}
    \includegraphics[width=5.3cm,height=3.4cm]{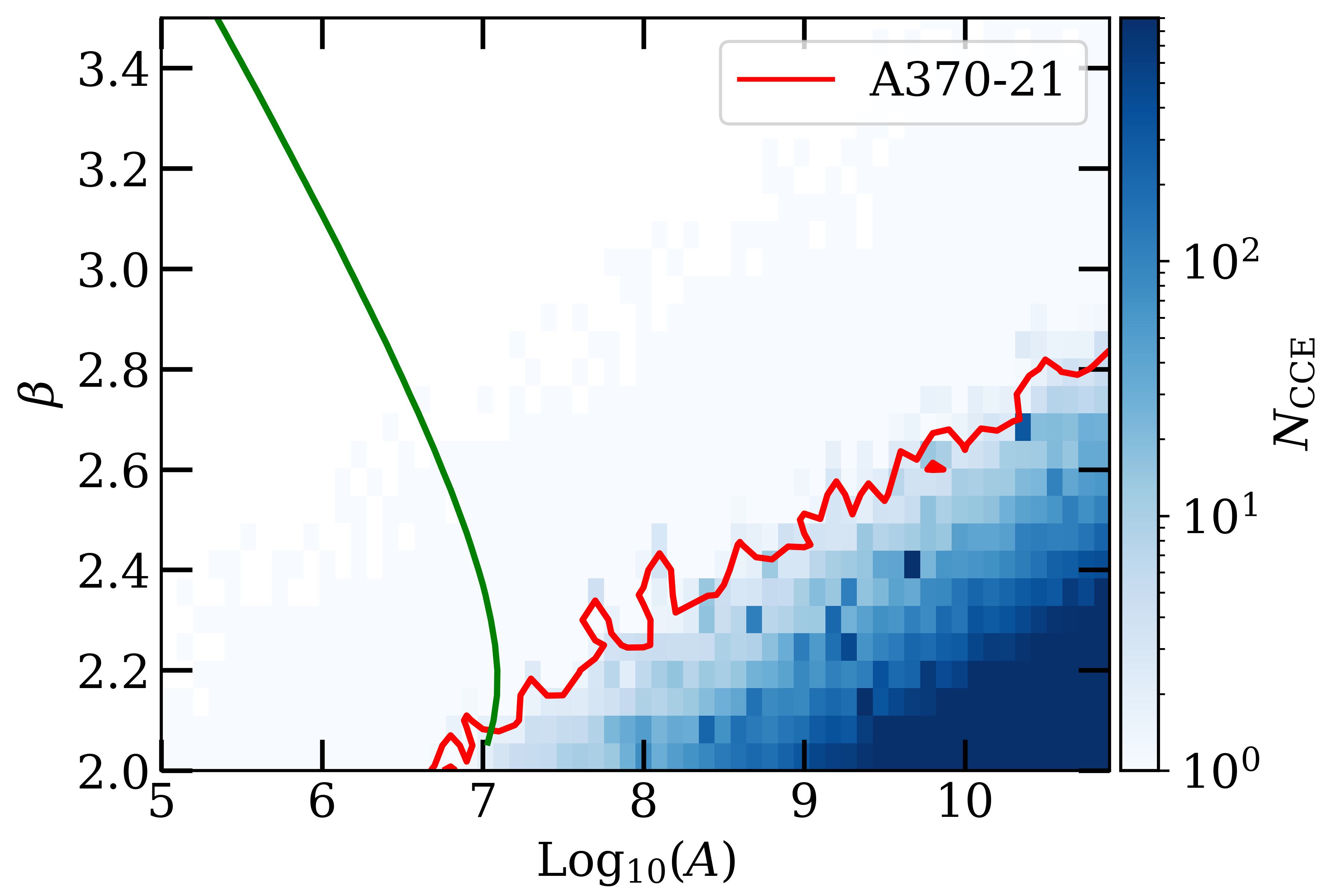}
    \includegraphics[width=5.3cm,height=3.4cm]{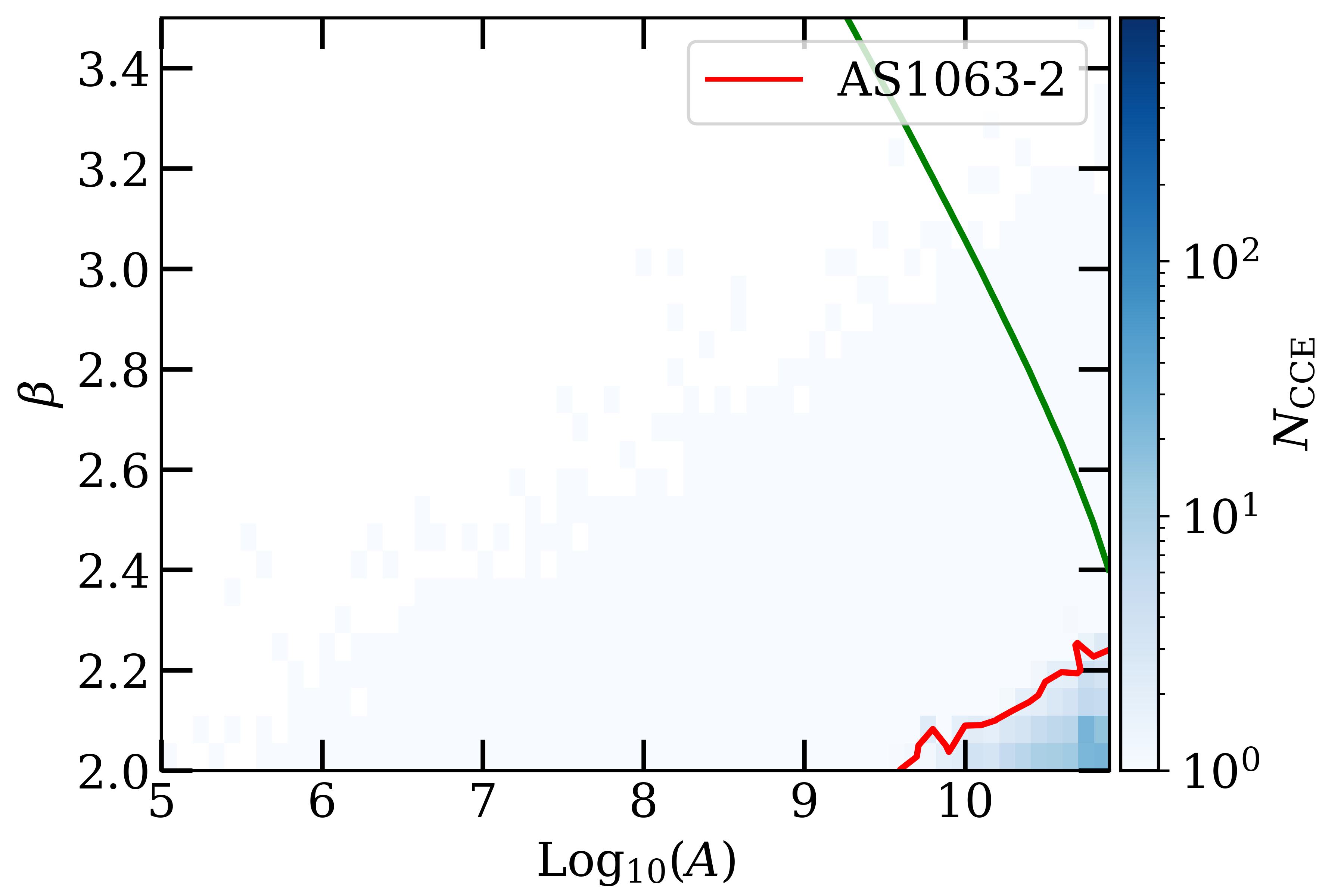}
    \includegraphics[width=5.3cm,height=3.4cm]{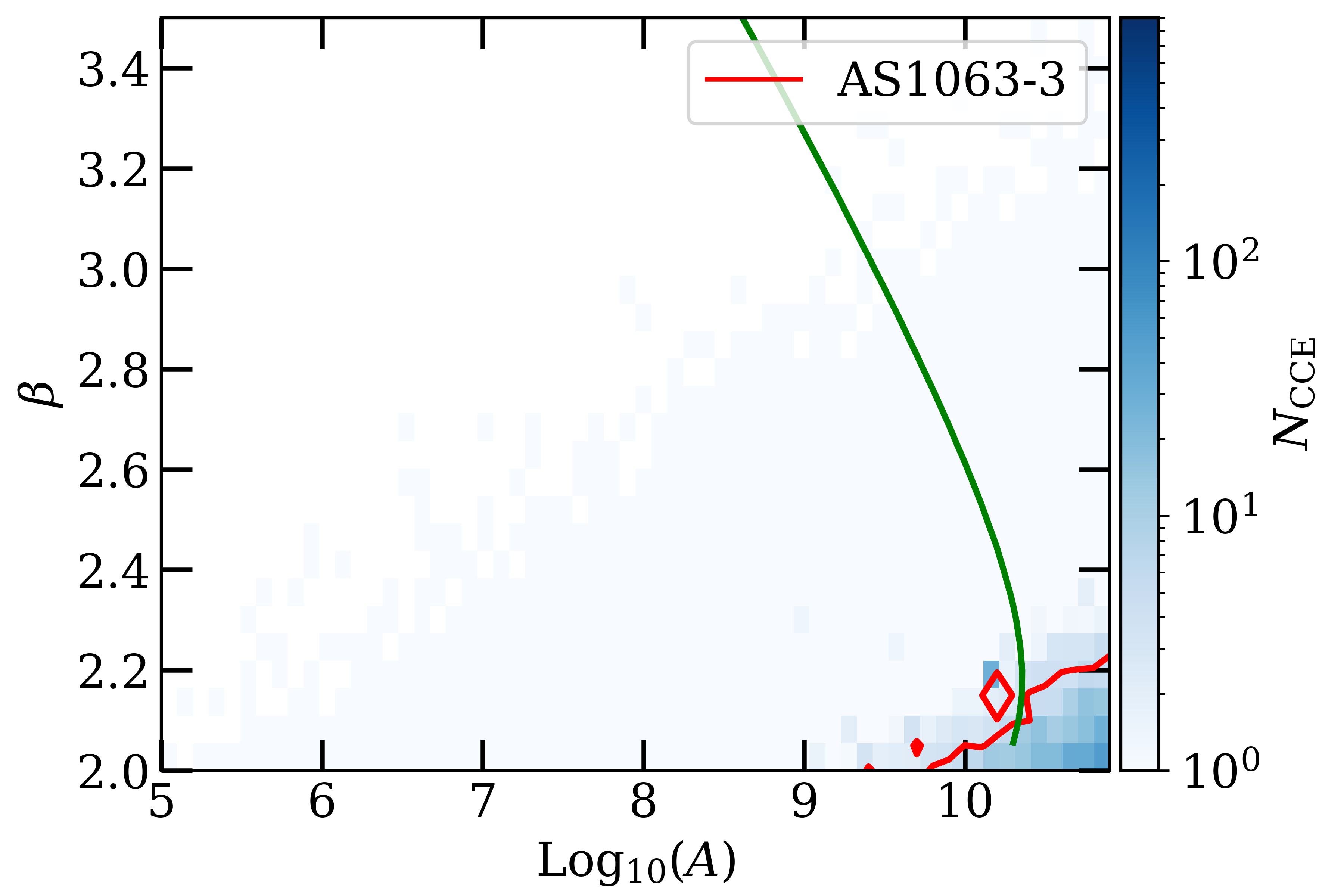}
    \includegraphics[width=5.3cm,height=3.4cm]{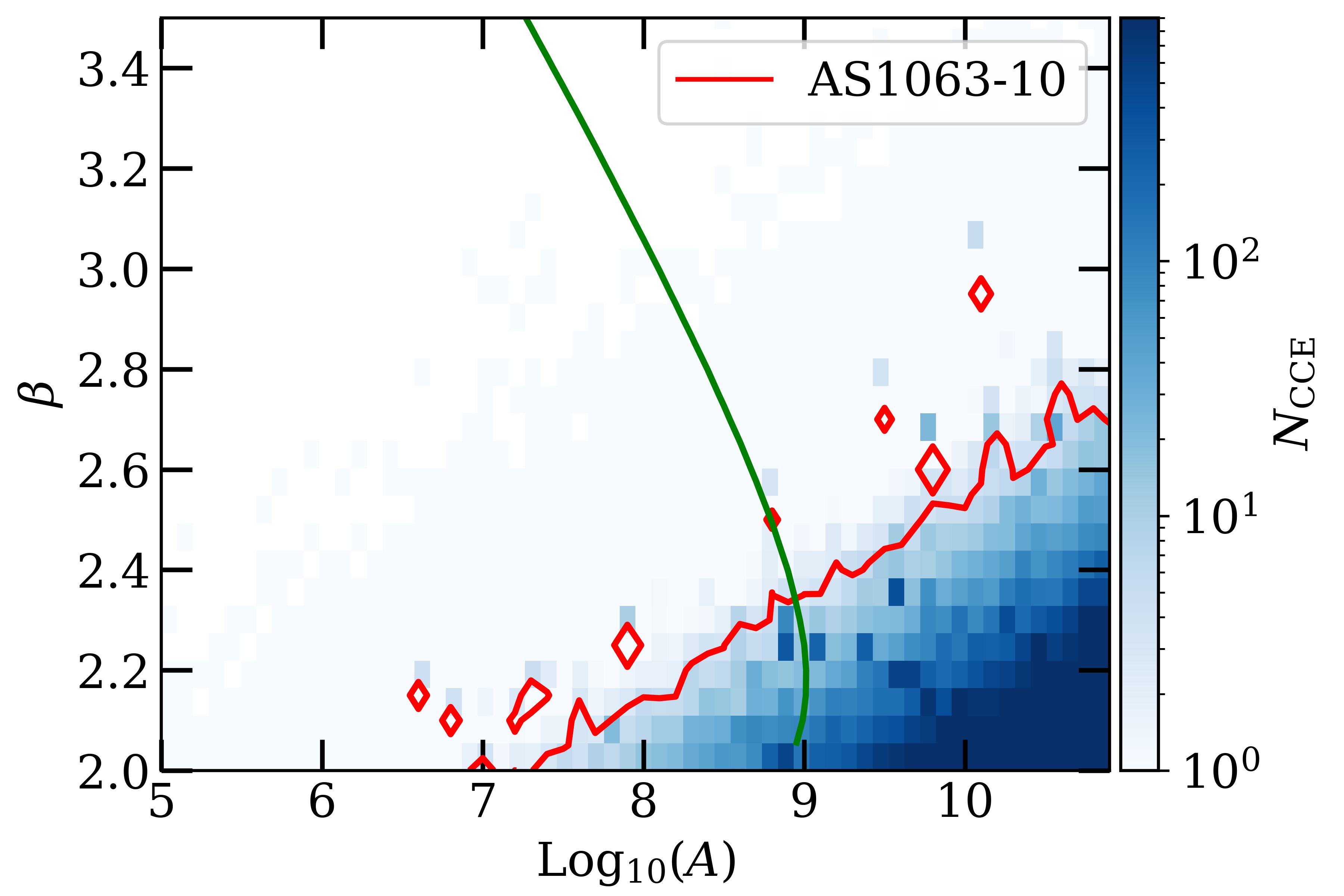}
    \includegraphics[width=5.3cm,height=3.4cm]{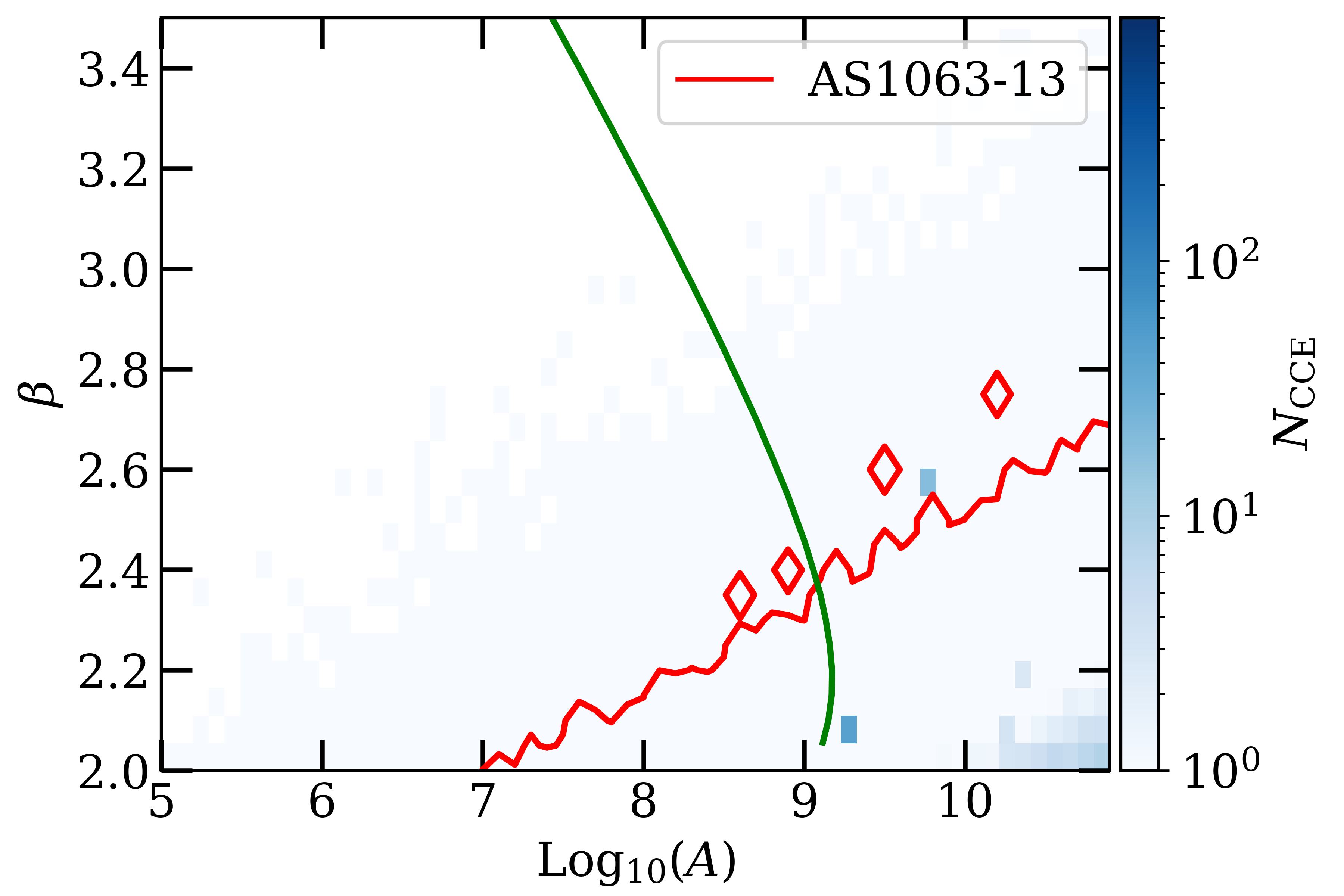}
    \includegraphics[width=5.3cm,height=3.4cm]{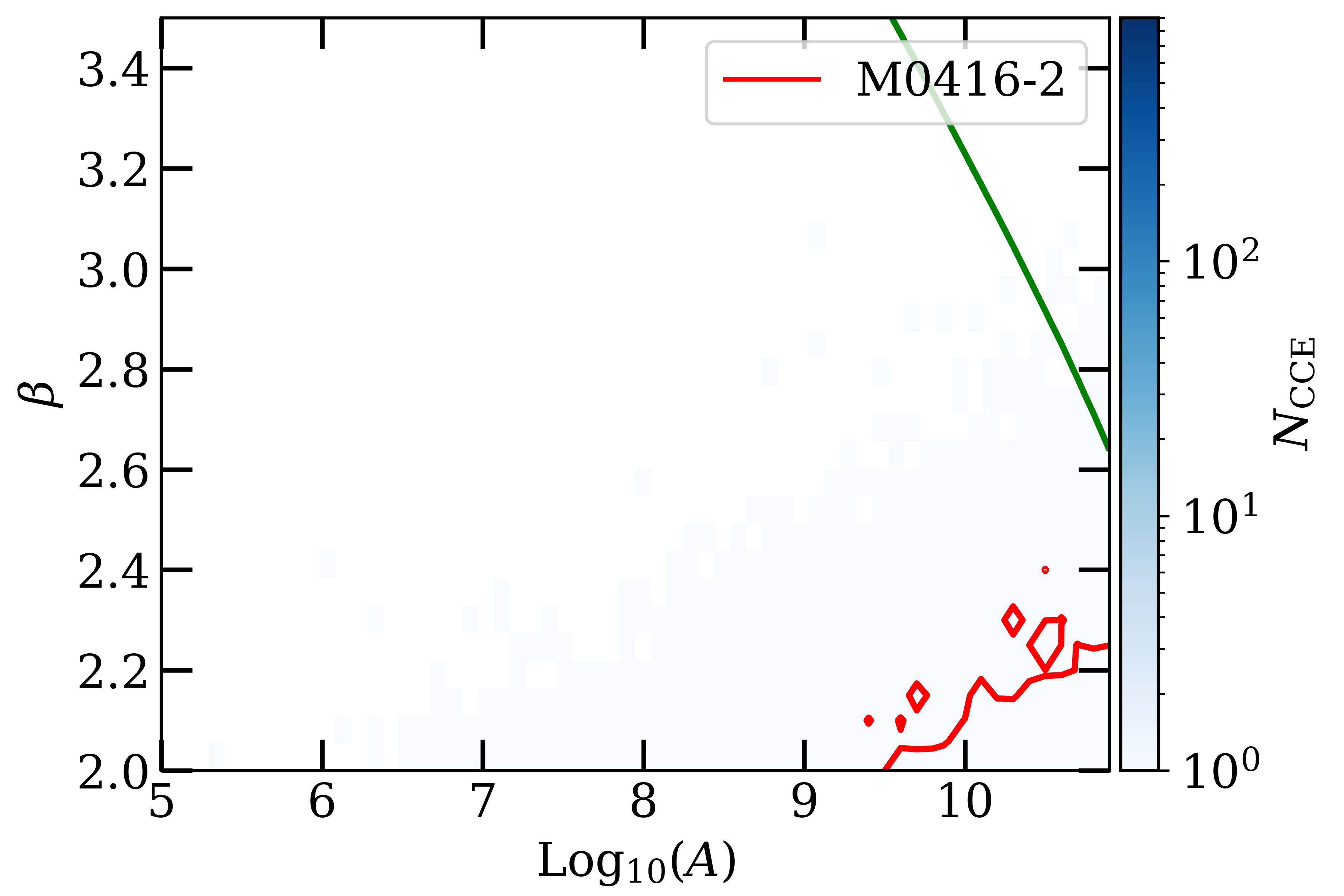}
    \includegraphics[width=5.3cm,height=3.4cm]{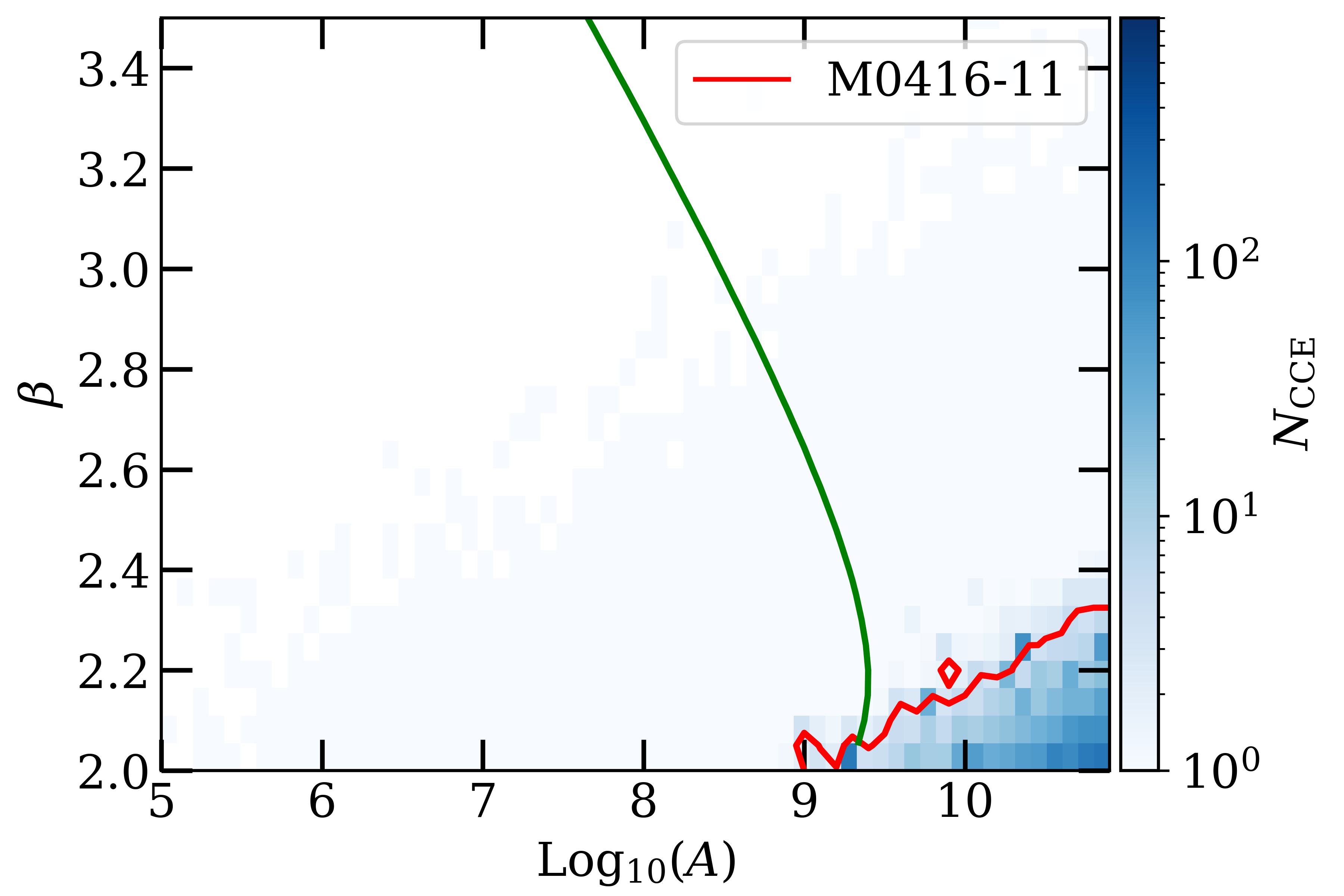}
    \includegraphics[width=5.3cm,height=3.4cm]{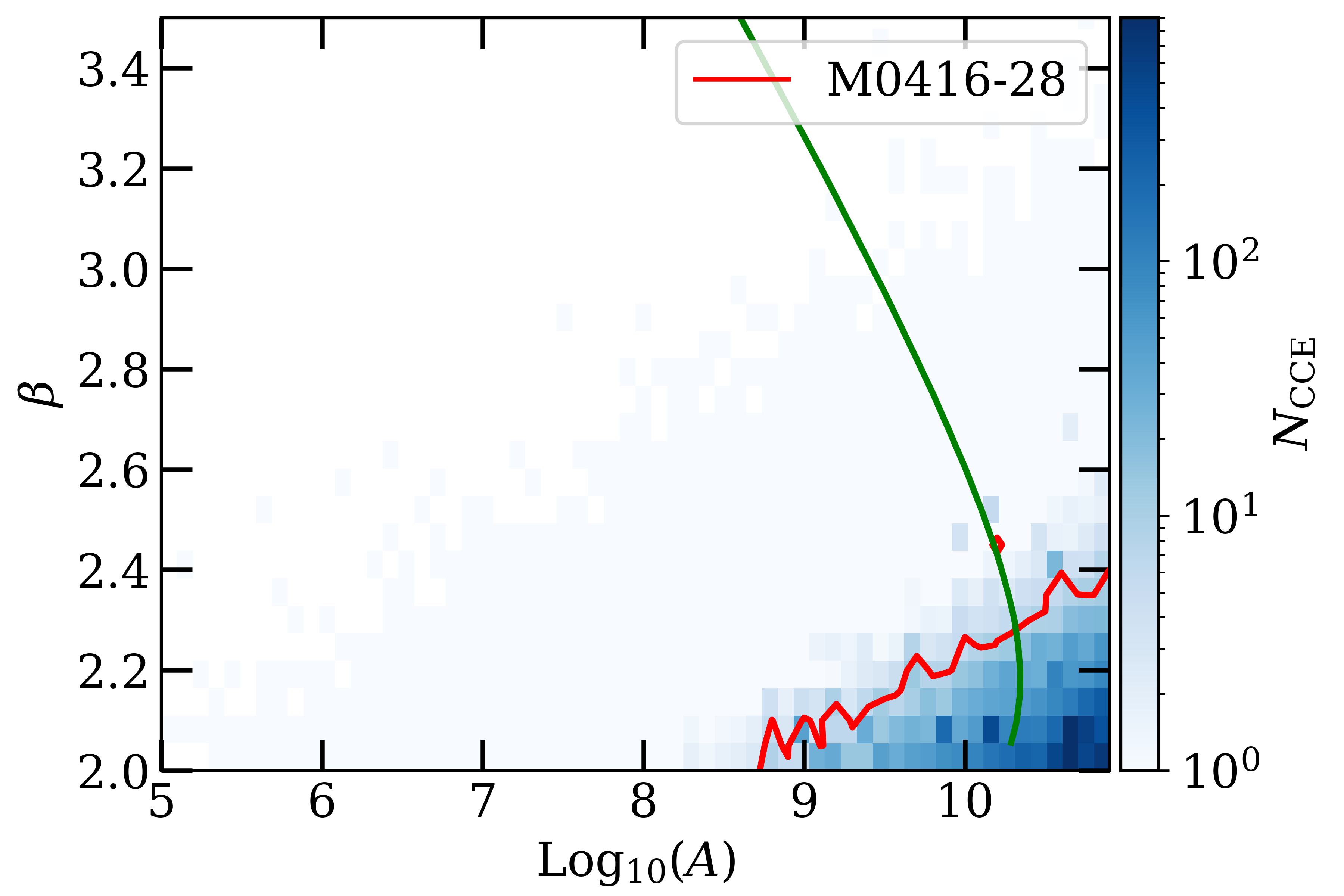}
    \includegraphics[width=5.3cm,height=3.4cm]{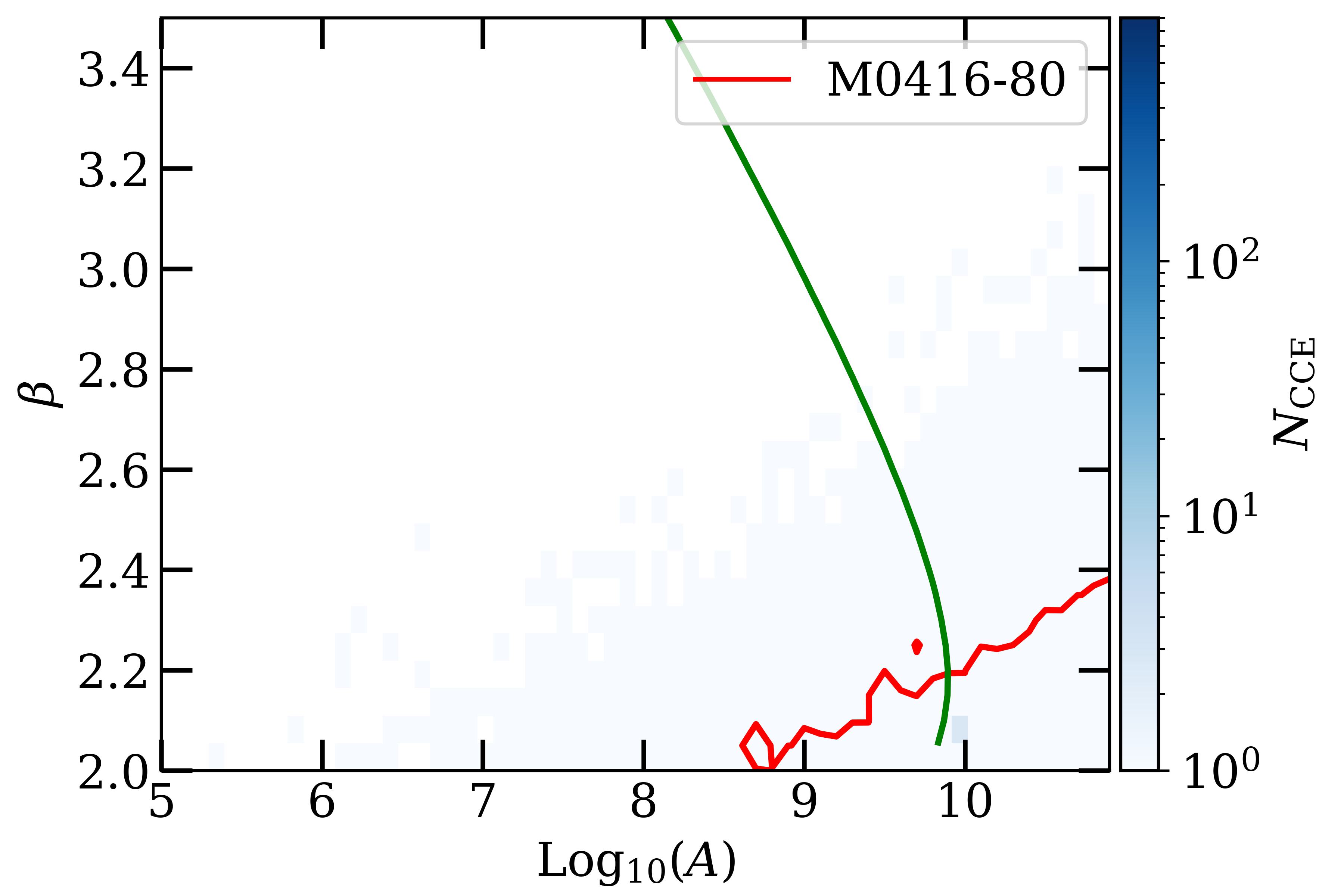}
    \includegraphics[width=5.3cm,height=3.4cm]{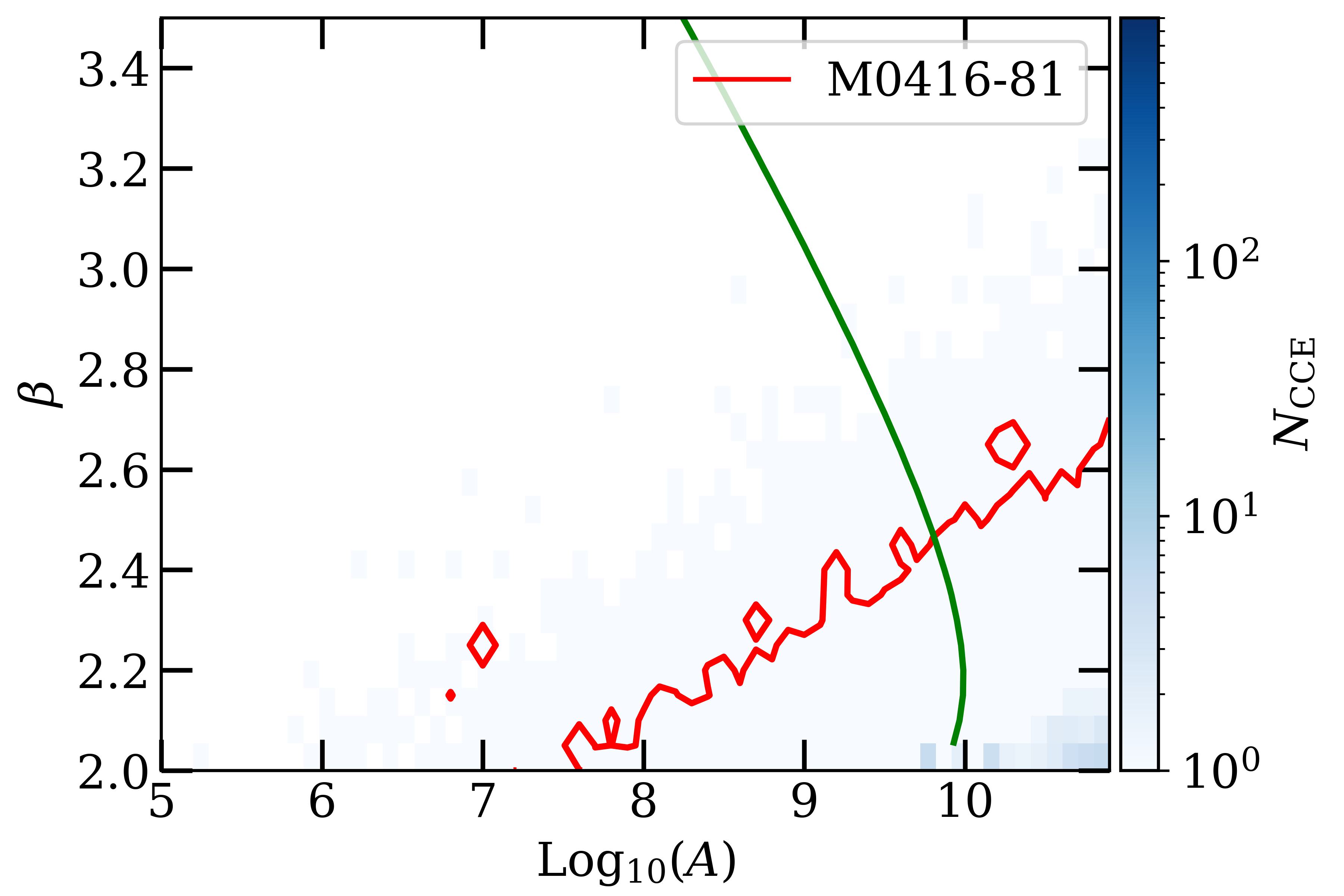}
    \includegraphics[width=5.3cm,height=3.4cm]{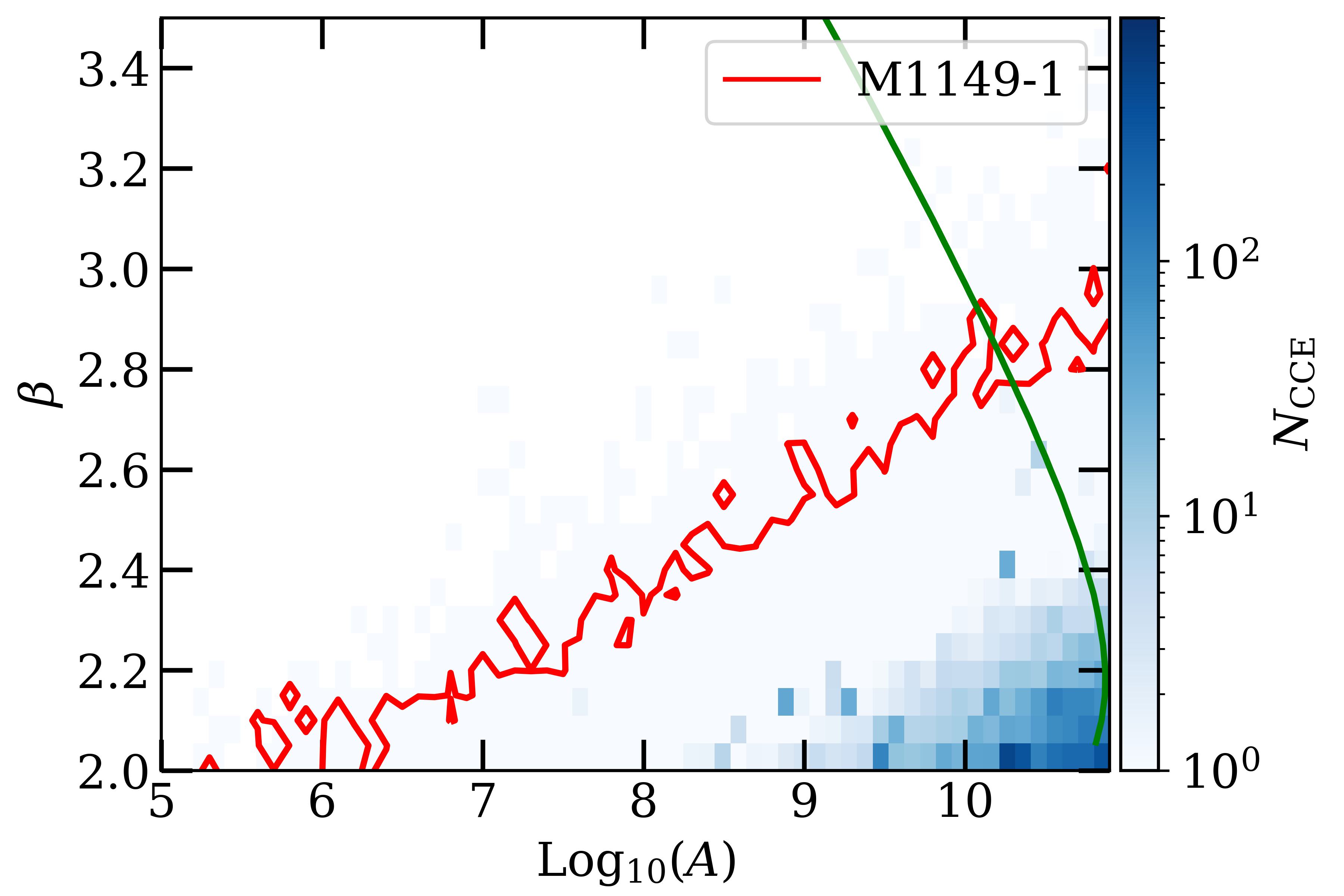}
    \includegraphics[width=5.3cm,height=3.4cm]{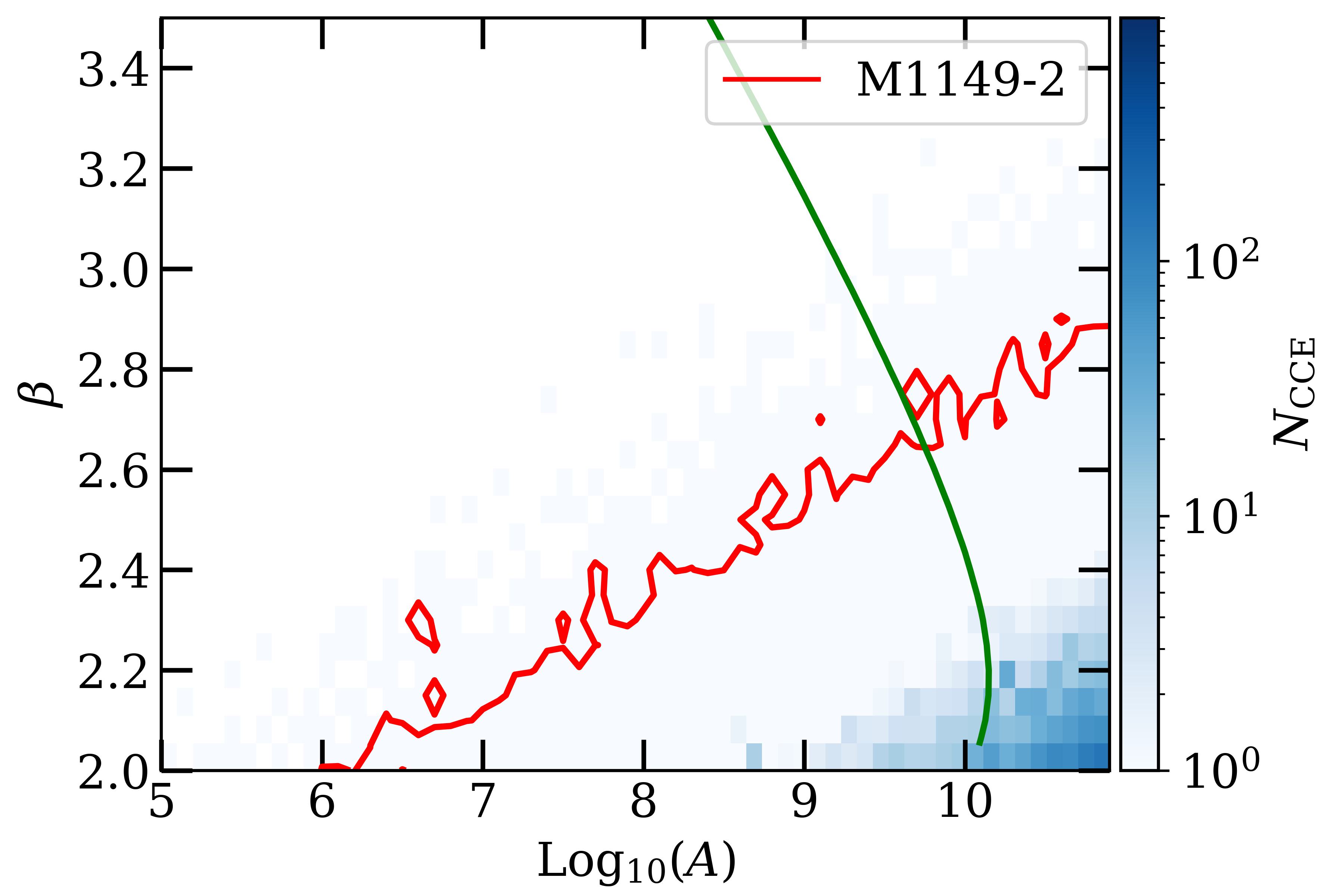}
    \caption{Number of CCEs in each arc assuming a power-law luminosity function. In each panel, the background color map represents the number of estimated CCEs given the value for~$(A, \beta)$ shown on the abscissa and ordinate axes. The red curve represents the number of observed CCEs. For arcs where we did not observe any CCEs, the red curve corresponds to~$N_{\rm CCE}=0.001$. In each panel, the green curves mark the~$(A, \beta)$ value that leads to a total luminosity equal to the observed lensing-corrected luminosity in F200LP, and the crossing point of green and red curves denotes the best-fit luminosity function slope~($\beta$) value for that particular arc.}
    \label{fig:ABeta}
\end{figure*}

\begin{figure}
    \centering
    \includegraphics[width=8.0cm,height=5.0cm]{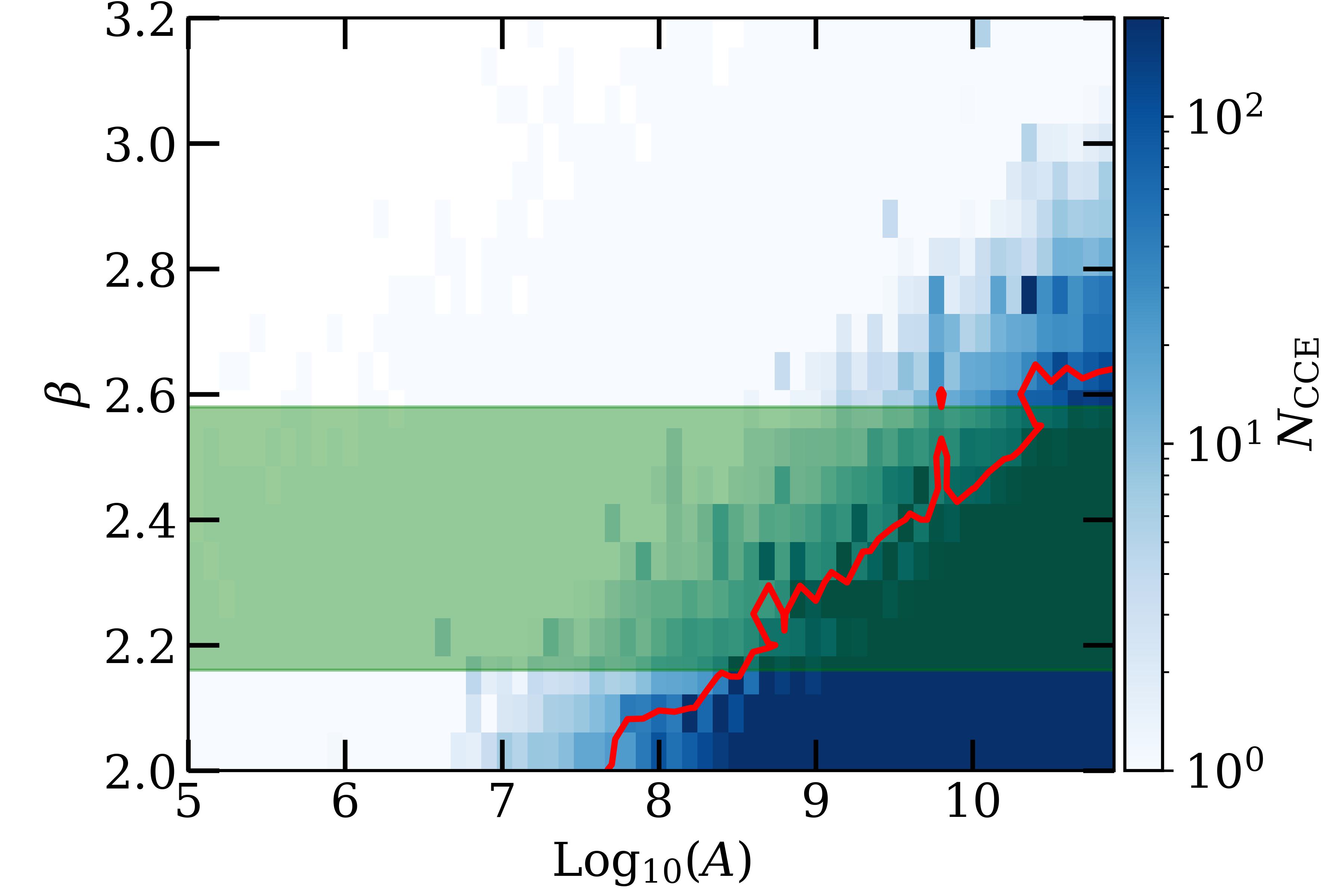}
    \caption{Total number of CCEs assuming a power-law luminosity function. The background color map represents the total number of expected CCEs during Flashlights observations in all of the lensed arcs considered in the current work. The red curve corresponds to~$N_{\rm CCE}=20$, i.e., the number of events observed in Flashlights. The shaded green region represents the best-fit $\beta$ value with $1\sigma$ range, i.e., $\beta=2.37\pm0.21$ (see Section~\ref{ssec:lum_fun} for more details).}
    \label{fig:AB_comb}
\end{figure}

\subsection{Constraining the rest-frame luminosity function of the stellar population}
\label{ssec:lum_fun}

In the analysis done above, we have simulated the stellar population of the sources and attempted to put initial constraints on the IMF based on the number of CCEs detected in the Flashlights survey. However, given the many (simplifying) assumptions, the simulated stellar population is quite model-dependent and, as a result, also the number of estimated CCEs. A more direct measurement based on the number of events would be to constrain the luminosity function~(in the F200LP filter in which the CCEs are observed). We assume that the observed F200LP luminosity function (per unit luminosity bin) describing the number of stars is given by a simple power law,
\begin{equation}
    dN = A\:L^{-\beta} dL,
    \label{eq:lf}
\end{equation}
where~$L$ is the unlensed F200LP luminosity and~$\beta$ is the slope of the unlensed observed luminosity function. Following this parametrization, we generated a stellar population corresponding to each arc as a function of~($A,\beta$) and predicted the number of CCEs for each such pair. We assume that a star can have F200LP luminosity values in the  range~$[0.01, 10^5]~{\rm L_\odot}$. We also have the F200LP luminosity measured in Flashlights for each arc, from which we can determine the viable range of~$A$ values. 

The results of our analysis for each lensed arc are shown in Figure~\ref{fig:ABeta}. In each panel, the background color map represents the expected number of CCEs in the particular system (mentioned in the corresponding legend), and the red line represents the observed number of CCEs given the~$(A, \beta)$ values. The green curve marks the values of~$(A, \beta)$ that lead to the total lensing-corrected observed luminosity of the arc, such that the intersection point of the two lines marks the corresponding best-fit $\beta$~value. The average best-fit $\beta$~value is~$2.37\pm0.21$. While estimating the~$\beta$~value, we used all arcs from Table~\ref{tab:arcs}. For arcs that do not lead to CCEs in Flashlights, we use intersection points assuming~$N_{\rm CCE}=0.001$ as shown in Figure~\ref{fig:ABeta}. If we only consider the arcs that actually led to CCEs in Flashlights, then we get a~$\beta$~value of~$2.23\pm0.12$, somewhat lower than earlier but within the error bars. We note that our estimations agree, within~$1\sigma$ uncertainties, with the~$\beta$~value of~$2.55_{-0.56}^{+0.72}$ reported by~\citet{2024A&A...689A.167D}, which is derived solely based on the distribution of microlensing events in the Dragon arc~(A370-Arc2). It should be noted, however, that although this measurement is free from many of the uncertainties that go into the stellar synthesis codes, it is still susceptible to the assumptions that go into our simplified CCE rate estimation~(see Sections~\ref{sec:rate} and~\ref{sec:uncertain} for more details). In Figure~\ref{fig:AB_comb}, we sum up the~$N_{\rm CCE}$ estimation for each arc and plot the~$N_{\rm CCE}=20$ curve (shown in red). Assuming that the average value of~$\log(A)$ for arcs in our sample lies in the~[8, 11] range, the corresponding $\beta$~value will lie in the~[2.0, 2.6] range, a more generous range compared to the above estimations.

\section{Uncertainties}
\label{sec:uncertain}
In the above analysis, we determined the number of CCEs expected during the Flashlights observations assuming two different IMFs for the source stellar population using a simple formalism that involves lens and stellar synthesis models with analytic calculations for the CCE rate. However, these results depend strongly on the simplifying assumptions made for our calculations and are thus prone to various sources of uncertainty. Some of these -- especially those originating from the stellar synthesis modeling -- are already encapsulated in the quoted errors, but other sources are not. In this section, we qualitatively describe various sources of uncertainties that can affect our results.

\subsection{Uncertainties in the macrolens model}
\label{ssec:uncertain_lens}
One important source of uncertainty that can largely affect the results and is not taken into account in our analysis is the uncertainty in the macromagnification~($\mu_{\rm m}$). Lensed arcs lie in high-magnification regions near the critical curve, where the uncertainties are also larger and can reach significant fractions~\citep{2015ApJ...801...44Z, 2017MNRAS.465.1030P, 2020MNRAS.494.4771R}. These uncertainties would directly influence the derived physical properties of the arc, and thus the estimated CCE rate. The uncertainty in the macromagnification primarily stems from our lack of knowledge about where the macrocritical curve intersects the arc, its orientation with respect to the arc, and the gradient of macromagnification in the arc.

For fold arcs, which are the most common, the macrocritical curve is expected to pass close to the midpoint of the arc, and we can determine the more accurate lens models just from visualization, reducing the uncertainty. That said, even if multiple lens models put the macrocritical curve at the same position in the middle of the lensed arc but have different orientations with respect to the arc, the estimated unlensed area for the model with a macrocritical curve perpendicular to the arc will be larger compared to the other models~(for example, see Figure~2 of \citealt{2023MNRAS.521.5224M}). For some fold arcs, determining the orientation of the macrocritical curve with respect to the arc may also be harder, as cluster lenses can have very complex critical curves. As we move farther away from the macrocritical curve, the macromagnification drops as $\mu\propto1/d$, where~$d$ is the distance from the macrocritical curve. The proportionality factor determines the variation of macromagnification near the macrocritical curve as well as the macrocaustic, yet it can differ by a factor of two between different lens models~\citep[e.g.,][]{2018ApJ...857...25D, 2024A&A...689A.167D}. These uncertainties will thus also be projected onto the number of expected CCEs, introducing a similar variation in the final CCE estimates. 

Detecting many CCEs in the same arc can also allow us to help determine preferred lens models as we expect to observe CCEs close to macrocritical curve positions. For example,~\citet{Keith24Prep} studies CCEs in the Spock arc~(M0416-Arc11) and claims that the lens models that lead to more than one critical curve crossing of Spock are favored over lens models with a single crossing, assuming that all observed transient events are CCEs.

\subsection{Uncertainties in the source stellar population}
\label{ssec:uncertain_ssps}
The parameters from \textsc{bagpipes} that go as input into \textsc{spisea} are~$(\psi, A_V, Z)$. Any uncertainty in the estimation of these parameters will affect the stellar population generated from \textsc{spisea}. Uncertainty in the above parameters can stem from either the modeling assumptions made by us, for example, regarding SFH, or priors, or under-the-hood assumptions used by the SED fitting code~\citep[e.g.,][]{2017ApJ...837..170L, 2023ApJ...944..141P}. In addition, there are strong degeneracies between various parameters, such as the age–dust–metallicity degeneracy~\citep[e.g.,][]{2001ApJ...550..212B}, which can affect the SFR~($\psi$) estimates. For example, \citet{Keith24Prep} considered a suite of parameter space and explored the difference in inferred CCE rate induced for the Spock arc and showed that by increasing $A_V = 0$ to $ = 0.6$, the CCE rate differs by a factor of $\sim$5 in F200LP. On the other hand, the effect of the choice of SFH is minimal where the inferred CCE rate in the same arc remains very similar across simulations using different SFH models. In our work, to account for the above uncertainties and degeneracies, as mentioned in Section~\ref{ssec:stel_pop}, we draw for each arc 10  realizations from the~$(\psi, A_V, Z)$ values within their $1\sigma$ ranges. However, this will only work well for a unimodal posterior distribution, and if any parameter shows multimodality, the uncertainties could further increase since the nominal $1\sigma$ range may cover less favored parameter values.

As we know, CCEs primarily originate from bright stars (whose evolution is uncertain and can be challenging to model) crossing the microcaustics. Hence, if the stellar evolution tracks in \textsc{spisea} are not properly accounting for certain effects -- for example, the effect of multiplicity on stellar evolution, which is expected to be common in O- and B-type stars~\citep[e.g.,][]{2023arXiv231101865M} -- the number of bright stars in the simulated stellar population would be affected, which in turn would affect the number of estimated CCEs. 

Another simplification made in our current work is that all stars have a fixed radius of~$20\:{\rm R_\odot}$. In reality, the radius of a star will depend on its luminosity and temperature, and to some extent on the stellar evolution model. Since the radius of the star affects the peak magnification it would attain (see Equation~\eqref{eq:mu_peak}), different stellar radii might thus also affect the resulting detection rate.

\subsection{Uncertainties related to the estimation of $R_{\rm CCE}$}
\label{ssec:uncertain_cce}
For a source star, given a macromagnification value, the rate of CCEs depends on CCE peak magnification, microlens density, and their mass function. For the CCE peak magnification estimation, following~\citet{2018PhRvD..97b3518O}, we have used an approximate formula that is correct at low microlens optical depths -- that is, outside the corrugated network. However, closer to the critical curve, this formula is expected to overestimate the peak magnification. To reduce the effect of this overestimation, we have limited the peak magnification to a conservative value of~$5\times10^3$, keeping in mind the saturation of peak magnification. Otherwise, the overestimation of the peak magnification would, in turn, increase the expected number of stars that should be observable during CCEs, ultimately leading to overestimation of the expected number of CCEs in Flashlights. Another possible source of uncertainty in CCE estimation is the number of CCEs per unit time. We estimate the number of CCEs by using the average number of microcaustics that a source star will cross per unit time, and doing so gives the maximum number of CCEs since it does not take into account the fact that often microcaustics will merge and decrease the number of CCEs; this effect will become more important as we move toward high magnification values in the lensed arc. This was also reflected by the comparison to explicit ray-tracing simulations seen in Figure~\ref{fig:compare}. At the same time, the merging of microcaustics can also affect the peak magnification, which our method will not capture. However, as discussed in Section~\ref{ssec:comp}, this effect is not expected to be significant in our current work, and a detailed investigation of this is left for future work. The number of microcaustics per unit time will also be affected by the choice of IMF used to deduce microlens density, as it affects the estimated mass-to-light ratio. For example, if instead of Kroupa we had assumed Salpeter IMF, it would have increased the mass-to-light ratio, leading to a higher microlens (and thus microcaustic) density roughly by a factor of two. That said, from Equation~\eqref{eq:mu_peak}, we can see that the corresponding peak magnification would have decreased.

Some of the uncertainties in the~$R_{\rm CCE}$ estimation can be eliminated if one uses a magnification probability density function~\citep[PDF;][]{2024PhRvD.110h3514K, 2024A&A...687A..81P} to draw random~$\mu$-values for each star given macromagnification and microlens density. However, this method comes with its own limitations. For example, one loses all temporal information, such as how long a microlensing peak will last, and one cannot study individual peaks. In addition, having two (or multiple) different methods to estimate the number of observable CCEs is beneficial to understanding various uncertainties and their effects on the number of expected CCEs. A detailed comparison of these different methods is an important exercise and will be presented in future work.

\subsection{Uncertainties in the nature of dark matter}
\label{ssec:uncertain_dm}
Another important factor that can introduce uncertainties in our analysis is the assumption that all dark matter is in the smooth form. If a considerable fraction of dark matter belongs to compact objects, then it will change the number of microcaustic crossing rates for the source star as well as the peak magnification that the star can achieve during the crossing. Recent works based on observed CCEs~\citep{2018PhRvD..97b3518O, 2024arXiv240408094W, 2025MNRAS.536.1579M} have suggested that compact dark with mass~$>10^{-6}{\rm M_\odot}$ can account for only~$\lesssim2\%$ of the dark matter. However, keeping in mind that the stellar mass density also only accounts for a percent level of the total surface mass density in clusters, even a percent level of compact dark matter can lead to considerable effects on the estimated number of CCEs. 

Similarly, the presence of subhalos affects the local distribution of macromagnification near macrocritical curves, affecting the number of CCEs \citep{2024ApJ...961..200W}. In the presence of subhalos, typically, we expect to see an increase in the number of CCEs far from the macrocritical curve compared to the no-subhalo case~\citep[e.g.,][]{2018ApJ...867...24D, 2024A&A...689A.167D}. Finally, if the nature of dark matter differs from the traditional cold dark matter model, such as wave dark matter~\citep[e.g.,][]{2020PhRvL.125k1102C, 2023NatAs...7..736A}, then it can also lead to an increment of CCEs in regions far from the macrocritical curve~\citep{2024A&A...689A.167D, 2025ApJ...978L...5B}. 

\subsection{Uncertainties in the nature of the transient}
\label{ssec:uncertain_src}
Above, we only focused on the uncertainties stemming from our lack of knowledge at various steps of our analysis while assuming that all of the detected transients are lensed stars. However, as the Flashlights survey only employs two long-pass filters, we cannot with certainty determine that all of these transients correspond to lensed stars. For example, \citet{2023MNRAS.521.5224M} discusses one of the transients detected in Flashlights in detail and shows that if it originated in the lensed source, then it is very likely to be a lensed star. Similarly,~\citet{2024arXiv240408571L} and \citet{2024A&A...689A.167D} discuss the contribution of luminous blue variables~(LBVs) to the transients detected in Abell~370. In addition, owing to the lack of color information in Flashlights, one cannot be certain that all observed transients reside in the lensed source, and any transients occurring in the cluster lens itself cannot be ruled out. Assuming that a certain fraction of the transients are not lensed stars would shift the observed values in Figure~\ref{fig:estimated_cce} toward lower values (in the direction of the Salpeter IMF). For example, A370-Arc21 gave rise to one transient in Flashlights~\citep{2023MNRAS.521.5224M}. However, based on our estimations, the probability of detecting such an event is very small as this arc itself is very faint and has a very small SFR, implying that this transient, if indeed a CCE, is a very rare one.

\section{Conclusion}
\label{sec:conclusions}
In this work, we have studied the ability of CCEs to put constraints on the IMF of bright, giant, and supergiant stars around cosmic noon with the Flashlights program data. Such estimates are crucial for understanding the formation and evolution of massive stars and their effects on galaxy evolution at cosmological distances where individual stars cannot otherwise be directly observed. 

We employed two distinct IMFs, namely Salpeter~($\alpha=2.35$) and top-heavy~($\alpha=1.0$), to generate stellar populations in the lensed galaxies and then estimated the expected number of CCEs in Flashlights observations. Based on our analysis, we find that the average IMF slope around cosmic noon is steeper than~$\alpha=1.0$ but shallower than the Salpeter~($\alpha=2.35$). As discussed above, however, our analysis is subject to various important uncertainties and is strongly model-dependent. Also, in our work, we have only probed two edge values for~$\alpha$. One could repeat the same analysis with more bins in~$\alpha$, as will be presented in an upcoming paper focusing on the Warhol arc~(M0416-Arc28).

In addition to constraining the IMF, we use the CCE rate to put direct constraints on the observed luminosity function in the arcs. Unlike the IMF measurement, this bypasses the uncertainties associated with the modeling of the source stellar population. We assume that the observed lensing-corrected F200LP luminosity of each source follows a simple power law, and by comparing our CCE estimates to the observed number of CCEs, we fit for the slope parameter~($\beta$). We find that the average is~$\beta=2.37\pm0.21$. Although this value is somewhat lower, it agrees with the results derived by~\citet{2024A&A...689A.167D} within~$1\sigma$. 

Keeping in mind the uncertainties involved, our current constraints on the IMF slope around cosmic noon should be considered only preliminary. Nonetheless, in this work, we present a simple (and efficient) method to constrain the IMF with CCE observations. Indeed, many more CCEs have been detected by now with JWST (increasing the CCE sample size and the quality of lens models), which could soon allow for more stringent constraints on the IMF throughout cosmic history. For example, a similar analysis of the Dragon arc (A370-Arc2), which recently led to more than 40 transients~\citep{2024arXiv240408045F} in JWST imaging, would allow us to put stringent constraints on the underlying IMF.

\begin{acknowledgements}
A.K.M. and A.Z. acknowledge support from Israel Science Foundation grant 86423, grant 2020750 from the United States-Israel Binational Science Foundation (BSF), grant 2109066  from the United States National Science Foundation (NSF), and the Ministry of Science \& Technology, Israel. S.K.L. acknowledges support from the Research Grants Council (RGC) of Hong Kong through the General Research Fund (GRF) 17302023. J.M.D. acknowledges support from project PID2022-138896NB-C51 (MCIU/AEI/MINECO/FEDER, UE) Ministerio de Ciencia, Investigación y Universidades. R.A.W. acknowledges support from NASA JWST Interdisciplinary Scientist grants NAG5-12460, NNX14AN10G, and 80NSSC18K0200 from GSFC. A.V.F. is grateful to the Christopher R. Redlich Fund and many other donors. The authors also thank the reviewer for useful comments.
\\
\\
This research was supported by NASA/HST grants GO-15936 and GO-16278 from the Space Telescope Science Institute, which is operated by the Association of Universities for Research in Astronomy, Inc., under NASA contract NAS5-26555. This work utilizes gravitational lensing models produced by PIs Bradac, Natarajan, \& Kneib (CATS), Merten \& Zitrin, Sharon, Williams, Keeton, Bernstein, Diego, and the GLAFIC group. This lens modeling was partially funded by the HST Frontier Fields program conducted by STScI. STScI is operated by the Association of Universities for Research in Astronomy, Inc., under NASA contract NAS 5-26555. The lens models were obtained from the Mikulski Archive for Space Telescopes (MAST). This research has made use of NASA’s Astrophysics Data System Bibliographic Services.
\\
\\
The work utilizes the following software packages:
\textsc{python}~(\url{https://www.python.org/}),
\textsc{astropy}~\citep{2018AJ....156..123A},
\textsc{bagpipes}~\citep{2018MNRAS.480.4379C},
\textsc{Jupyter Notebook}~\citep{2016ppap.book...87K},
\textsc{Matplotlib}~\citep{2007CSE.....9...90H},
\textsc{NumPy}~\citep{2020Natur.585..357H}, and
\textsc{spisea}~\citep{2020AJ....160..143H}.
\end{acknowledgements}

\bibliographystyle{aa}
\bibliography{reference}

\onecolumn		
\begin{appendix}
\section{Sample of lensed arcs and their properties}

\begin{longtable}{cccccccccc}
    \caption{Strongly lensed arcs used in our current work observed in HFF clusters. Columns (1), (2), (3), and (4) represent the image cluster name ID, R.A.,  Dec., and source redshift of the lensed arc, respectively. Column~(5) gives the ICL surface density near each lensed arc. Columns~(6),~(7), and (8) provide the star-formation rate~(SFR), metallicity~($Z$), and dust extinction~($A_V$) for each arc estimated using \textsc{bagpipes}. Columns~(9) and (10) list the average number of CCEs per visit in each arc for the top-heavy~($\alpha=1.00$) and Salpeter~($\alpha=2.35$) IMFs, respectively. For Abell~370, Abell~S1063, MACS0416, and MACS1149, the coordinates and arc redshifts are taken from~\citet{2019MNRAS.485.3738L}, \citet{2016A&A...587A..80C}, \citet{2021AA...646A..83R}, and \citet{2021ApJ...919...54Z}, respectively, and references therein.} \\
    \hline
    \hline
	ID &   R.A.    &  Dec.    & $z_s$ &   $\Sigma_{\rm ICL}$    & SFR                    & $Z$                    &  $A_v$                 & $N_{\rm CCE} (\alpha=1.00)$ & $N_{\rm CCE} (\alpha=2.35)$ \\
	&  (J2000)  &  (J2000) &       &   ${\rm M_\odot/pc^2}$  & (${\rm M_\odot/yr}$)   & (in $Z_\odot$)         & (in mag)               &                             &                             \\ 
	(1) &  (2)      &  (3)     & (4)   &   (5)                   & (6)                    &  (7)                   & (8)                    & (9)                         &  (10)                       \\
	\hline
    \endfirsthead
    \caption{continued.}\\
    \hline\hline
	ID &   R.A.    &  Dec.    & $z_s$ &   $\Sigma_{\rm ICL}$    & SFR                    & $Z$                    &  $A_v$                 & $N_{\rm CCE} (\alpha=1.00)$ & $N_{\rm CCE} (\alpha=2.35)$ \\
	&  (J2000)  &  (J2000) &       &   ${\rm M_\odot/pc^2}$  & (${\rm M_\odot/yr}$)   & (in $Z_\odot$)         & (in mag)               &                             &                             \\ 
	(1) &  (2)      &  (3)     & (4)   &   (5)                   & (6)                    &  (7)                   & (8)                    & (9)                         &  (10)                       \\
    \hline
    \endhead
    \hline
    \endfoot
    &           &          &       &                         &   Abell 370            &                        &                        &                             &                             \\
	1.1 & 39.96704  & -1.57691 & 0.8041& $12.03^{+3.72}_{-2.99}$ & $0.15^{+0.06}_{-0.04}$ & $0.88^{+0.04}_{-0.04}$ & $1.11^{+0.06}_{-0.04}$ & $4.27\pm1.10$               & $1.39\pm0.87$               \\
	1.2 & 39.97627  & -1.57605 &       &                         &                        &                        &                        &                             &                             \\
	1.3 & 39.96869  & -1.57661 &       &                         &                        &                        &                        &                             &                             \\
	\\
	2.1 & 39.97382  & -1.58422 & 0.7251& $33.41^{+9.73}_{-11.96}$& $1.05^{+0.33}_{-0.19}$ & $0.38^{+0.31}_{-0.24}$ & $0.97^{+0.98}_{-0.13}$ & $8.63\pm5.60$               & $4.25\pm1.60$               \\
	2.2 & 39.97100  & -1.58504 &       &                         &                        &                        &                        &                             &                             \\
	2.3 & 39.96872  & -1.58450 &       &                         &                        &                        &                        &                             &                             \\
	2.4 & 39.96939  & -1.58473 &       &                         &                        &                        &                        &                             &                             \\
	2.5 & 39.96963  & -1.58485 &       &                         &                        &                        &                        &                             &                             \\
	\\
	3.1 & 39.96565  & -1.56685 & 1.9553& $0.24^{+0.14}_{-0.23}$  & $0.06^{+0.08}_{-0.04}$ & $0.37^{+0.34}_{-0.28}$ & $0.09^{+0.10}_{-0.06}$ & $0.00\pm0.00$               & $0.0\pm0.0$                 \\
	3.2 & 39.96852  & -1.56579 &       &                         &                        &                        &                        &                             &                             \\
	3.3 & 39.97892  & -1.56746 &       &                         &                        &                        &                        &                             &                             \\
	\\
	5.1 & 39.97347  & -1.58904 & 1.2775& $0.19^{+0.18}_{-0.19}$  & $0.03^{+0.02}_{-0.01}$ & $0.50^{+0.34}_{-0.33}$ & $0.33^{+0.30}_{-0.21}$ & $0.01\pm0.00$               & $0.00\pm0.00$               \\
	5.2 & 39.97057  & -1.58919 &       &                         &                        &                        &                        &                             &                             \\
	5.3 & 39.96947  & -1.58909 &       &                         &                        &                        &                        &                             &                             \\
	5.4 & 39.96858  & -1.58900 &       &                         &                        &                        &                        &                             &                             \\
	\\
	7.1 & 39.96978  & -1.58042 & 2.7512& $21.70^{+6.77}_{-9.36}$ & $0.35^{+1.48}_{-0.35}$ & $0.77^{+0.17}_{-0.34}$ & $0.69^{+0.20}_{-0.35}$ & $0.00\pm0.00$               & $0.00\pm0.00$               \\
	7.2 & 39.96988  & -1.58076 &       &                         &                        &                        &                        &                             &                             \\
	7.3 & 39.96881  & -1.58563 &       &                         &                        &                        &                        &                             &                             \\
	7.4 & 39.98656  & -1.57756 &       &                         &                        &                        &                        &                             &                             \\
	7.5 & 39.96153  & -1.58000 &       &                         &                        &                        &                        &                             &                             \\
	\\
	21.1& 39.96673  & -1.58469 & 1.2567& $4.73^{+1.54}_{-1.50}$  & $<0.01^{+0.01}_{-0.01}$ & $0.06^{+0.23}_{-0.04}$ & $0.12^{+0.17}_{-0.09}$ & $0.00\pm0.00$               & $0.00\pm0.06$               \\
	21.2& 39.96725  & -1.58496 &       &                         &                        &                        &                        &                             &                             \\
	21.3& 39.98153  & -1.58140 &       &                         &                        &                        &                        &                             &                             \\
	\hline
	&           &          &       &                         &   Abell S1063          &                        &                        &                             &                             \\ 
	2.1 & 342.19588 & -44.52895& 1.2278& $1.47^{+2.46}_{-1.43}$  & $0.94^{+0.52}_{-0.23}$ & $0.97^{+0.02}_{-0.05}$ & $1.29^{+0.09}_{-0.09}$ & $0.89\pm0.81$               & $0.15\pm0.10$               \\
	2.2 & 342.19450 & -44.52698&       &                         &                        &                        &                        &                             &                             \\
	2.3 & 342.18642 & -44.52116&       &                         &                        &                        &                        &                             &                             \\
	2.4 & 342.19520 & -44.52786&       &                         &                        &                        &                        &                             &                             \\
	\\
	3.1 & 342.19271 & -44.53119& 1.2592& $0.99^{+2.77}_{-0.96}$  & $0.23^{+0.15}_{-0.11}$ & $0.83^{+0.09}_{-0.12}$ & $1.11^{+0.08}_{-0.09}$ & $0.11\pm0.06$               & $0.09\pm0.40$               \\
	3.2 & 342.19213 & -44.52983&       &                         &                        &                        &                        &                             &                             \\
	3.3 & 342.17983 & -44.52156&       &                         &                        &                        &                        &                             &                             \\
	\\
	10.1& 342.19023 & -44.52976& 0.7287& $15.80^{+6.05}_{-6.90}$ & $0.03^{+0.02}_{-0.02}$ & $0.16^{+0.04}_{-0.20}$ & $0.23^{+0.25}_{-0.16}$ & $1.38\pm0.40$               & $0.54\pm0.00$               \\
	10.2& 342.18995 & -44.52884&       &                         &                        &                        &                        &                             &                             \\
	10.3& 342.19491 & -44.52547&       &                         &                        &                        &                        &                             &                             \\
	\\
	13.1& 342.19369 & -44.53014& 1.2583& $0.06^{+1.72}_{-0.04}$  & $0.01^{+0.01}_{-0.01}$ & $0.53^{+0.32}_{-0.40}$ & $0.12^{+0.14}_{-0.09}$ & $0.00\pm0.00$               & $0.00\pm0.00$               \\
	13.2& 342.19331 & -44.52942&       &                         &                        &                        &                        &                             &                             \\
	\hline
	&           &          &       &                         &  Macs~0416             &                        &                        &                             &                             \\
	2.1 & 64.04116  & -24.06188& 1.8925& $0.16^{+1.84}_{-0.11}$  & $7.46^{+5.13}_{-2.09}$ & $0.64^{+0.24}_{-0.28}$ & $0.93^{+0.27}_{-0.35}$ & $0.00\pm0.00$               & $0.00\pm0.00$               \\
	2.2 & 64.04300  & -24.06303&       &                         &                        &                        &                        &                             &                             \\
	2.3 & 64.04745  & -24.06885&       &                         &                        &                        &                        &                             &                             \\
	\\
	11.1& 64.03920  & -24.07036& 1.0060& $1.80^{+6.10}_{-1.72}$  & $0.01^{+0.01}_{-0.01}$ & $0.12^{+0.01}_{-0.01}$ & $0.12^{+0.10}_{-0.07}$ & $0.00\pm0.00$               & $0.00\pm0.00$               \\
	11.2& 64.03833  & -24.06975&       &                         &                        &                        &                        &                             &                             \\
	11.3& 64.03425  & -24.06601&       &                         &                        &                        &                        &                             &                             \\
	\\
	28.1& 64.03645  & -24.06702& 0.9394&$30.00^{+14.23}_{-13.35}$& $0.29^{+0.04}_{-0.03}$ & $0.12^{+0.01}_{-0.01}$ & $1.20^{+0.09}_{-0.08}$ & $1.54\pm0.49$               & $0.15\pm0.06$               \\
	28.2& 64.03687  & -24.06749&       &                         &                        &                        &                        &                             &                             \\
	28.3& 64.04091  & -24.07115&       &                         &                        &                        &                        &                             &                             \\
	\\
	80.1& 64.04108  & -24.06308& 2.2430& $0.26^{+0.71}_{-0.18}$  & $0.08^{+0.07}_{-0.03}$ & $0.56^{+0.30}_{-0.28}$ & $0.34^{+0.24}_{-0.22}$ & $0.00\pm0.00$               & $0.00\pm0.00$               \\
	80.2& 64.04091  & -24.06296&       &                         &                        &                        &                        &                             &                             \\
	\\
	81.1& 64.03683  & -24.06634& 2.0910& $5.13^{+5.38}_{-4.98}$  & $1.88^{+1.50}_{-1.00}$ & $0.66^{+0.24}_{-0.31}$ & $0.75^{+0.35}_{-0.36}$ & $0.00\pm0.00$               & $0.00\pm0.00$               \\
	81.2& 64.03658  & -24.06612&       &                         &                        &                        &                        &                             &                             \\
	\hline
	&           &          &       &                         & Macs~1149              &                        &                        &                             &                             \\
	1.1 & 177.39700 & 22.39600 & 1.4880&$28.43^{+12.27}_{-11.48}$& $0.01^{+0.01}_{-0.01}$ & $0.97^{+0.02}_{-0.04}$ & $0.01^{+0.02}_{-0.01}$ & $0.19\pm0.14$               & $0.01\pm0.01$               \\
	1.2 & 177.39942 & 22.39743 &       &                         &                        &                        &                        &                             &                             \\
	1.3 & 177.40342 & 22.40243 &       &                         &                        &                        &                        &                             &                             \\
	\\            
	2.1 & 177.40242 & 22.38975 & 1.8910& $3.00^{+1.16}_{-1.19}$  & $0.10^{+0.03}_{-0.03}$ & $0.22^{+0.07}_{-0.05}$ & $0.37^{+0.12}_{-0.13}$ & $0.44\pm0.22$               & $0.00\pm0.00$               \\
	2.2 & 177.40604 & 22.39247 &       &                         &                        &                        &                        &                             &                             \\
	2.3 & 177.40658 & 22.39288 &       &                         &                        &                        &                        &                             &                             \\
    \label{tab:arcs}
\end{longtable}
\end{appendix}

\twocolumn

\end{document}